\documentclass[reprint, amsmath,amssymb, aps, pre, floatfix, 10pt]{revtex4-2}
\usepackage{algorithm}
\usepackage{algpseudocode}
\usepackage{dcolumn}% Align table columns on decimal point
\usepackage{bm}% bold math
\usepackage{graphicx}

\newcommand{\lss}{\boldsymbol{\ell}_{s}}
\newcommand{\ltt}{\boldsymbol{\ell}_{t}}
\newcommand{\cf}{{\it cf.}~}

\begin{document}
\title{Empirical Discovery of Multi-Scale Transfer of Information in Dynamical Systems}

\author{Christopher W. Curtis}
\affiliation{Department of Mathematics and Statistics \\ San Diego State University, 5500 Campanile Dr., San Diego, CA, 92182}

\author{Erik M. Bollt}
\affiliation{Department of Electrical and Computer Engineering, Clarkson University, 8 Clarkson Avenue, Potsdam, New York 13699, USA}
\affiliation{Clarkson Center for Complex Systems Science, Clarkson University, 8 Clarkson Avenue, Potsdam, New York 13699, USA}

\begin{abstract}
In this work, we quantify the time scales and information flow associated with multiscale energy transfer in a weakly turbulent system.  This is done through a greedy optimization algorithm which finds the maximum conditional-mutual information across lagged embeddings of time series localized by wavenumber.  For our chosen weakly turbulent system, the algorithm finds asymmetries in the information flow across wavenumbers, reflecting what are typically described as forward and inverse cascades.  However, our approach goes beyond typical heuristic arguments and provides quantitative insight into the intricate multi-wave mixing dynamics necessary to maintain the steady statistical state characterizing weak turbulence.  Our work then provides a novel, detailed, and fully nonlinear statistical analysis of a weakly turbulent system.  The flexibility of our approach points to broader applicability in real-world data coming from chaotic or turbulent dynamical systems.  
\end{abstract}
\date{\today}

\maketitle
\section{Introduction}
The question of causality, or perhaps more broadly information flow and coupling, in time series is a central one.  By addressing the question in linear time series coming from econometric data \cite{granger}, Clive Granger famously won a Nobel prize in 2003.  Building off of this ground-breaking work, methods using information theory to determine informative couplings between variables in nonlinear time series have been developed; see in particular \cite{schreiber} which introduced the metric of {\it transfer entropy} (TE).  TE is equivalent to also computing the {\it conditional-mutual information} (CMI), which is a central quantitative tool for determining causal relationships in time series of multidimensional processes.  See \cite{palus} for a more recent discussion and elaboration of the work in \cite{schreiber}.  Furthermore, we have shown, building from the TE/CMI identity, that conditioning on tertiary effects by what we called causation entropy (CE) \cite{bollt, bollt2,sun2014causation,sun2014identifying}, allows for an effective means of identifying causal chains across large numbers of measured variables.  This was implemented in an algorithm that we called optimal causation entropy (oCSE) that allows for accurately generating networks of information flow among multiple time series.  Readily available and dedicated software libraries, such as IDTxl \cite{idtxl, idtxl2}, now make the generation of these networks increasingly straightforward.  

However, the question of how information flows can describe processes in fluid mechanics and related subjects like nonlinear wave equations is still a relatively new topic.  Preliminary results within the last decade in \cite{e16031272} led in part to the recent work in \cite{lozano, lozano2, Arranz_Lozano-Durán_2024} which shows how TE and its generalizations can provide a very sophisticated understanding of the measurement of energy cascades in fluid turbulence.  See also the review in \cite{CAMPSVALLS20231}, which provides a lengthy exploration of the development of this subject from a physics perspective.  Likewise, studies of information flow in chaotic and turbulent dynamical systems have appeared with regard to modeling error quantification and generating corresponding fluctuation-dissipation models; see \cite{chen} and related work.  Preliminary work exploring how information theory helps describe atmospheric and ionospheric dynamics has also appeared in \cite{materassi1, materassi2}. Nevertheless, much remains to be explored in this subject.

Adding then to this area, in this paper, we study how information theory is able to track multiscale energy transfer in a version of the Majda-McLaughlin-Tabak (MMT) model \cite{majda1, majda2}.  The version we pick is distinguished by the absence of the formation of coherent structures that otherwise violate the assumptions typifying WWT regimes.  This model is interesting since despite its relative simplicity of being only a 1+1 dimensional nonlinear dispersive-wave equation, it is known to exhibit weak-wave turbulence (WWT) \cite{majda1, zakharov}, and both forward and inverse cascades are present.  Moreover, tracking information flow in WWT is a nontrivial task as recent results from \cite{RUMPF2005188, PhysRevLett.103.074502, PhysRevE.95.062225, lvov, hrabski, Simonis_Hrabski_Pan_2024} have shown.  While WWT is characterized by a statistically stationary distribution, since energy is transported via multi-wave mixing \cite{PhysRevE.95.062225}, straightforward descriptions of energy and particle number flow across scales, in contrast to classical fluids, are not available.  To analyze these complex phenomena then, we modify the IDTxl library to better track the most information rich couplings in multidimensional nonlinear time series.  Aside from changing the model design and hypothesis testing process, our modifications also utilize a particular sorting method to address what would otherwise lead to a combinatorial explosion of options.  We label our algorithm the sorted IDTxl (sIDTxl) method and establish its validity through testing on a coupled Lorenz--R\"{o}ssler system, which is a popular nonlinear and chaotic benchmark \cite{faes}.  

While we directly apply sIDTxl to the dimensions of the coupled Lorenz--R\"{o}ssler system, in the MMT model we apply it to the wavenumber separated fluctuations around the stationary equilibrium distribution.  These fluctuations are shown to be typified by having nearly Gaussian distributions with nonlinearity manifesting as significant skewness with a positive bias.  In line with \cite{PhysRevE.95.062225, hrabski, lvov}, our results from sIDTxl show that information flow occurs across a plurality of wavenumbers and time scales that generally eschews a simple characterization via direct or inverse cascades.  Insofar as such characterizations are possible, we see that this comes from an asymmetry in which forward `energy' transfer represents the larger portion of information flow across wavenumbers.  This asymmetry is most pronounced when in sIDTxl we limit the timescale of lagged couplings between fluctuations to one full nonlinear `turnover' time in the MMT equation.  Looking out over two turnover times creates a more symmetric information exchange landscape, though we still see a bias towards direct cascades of energy. While overall our results generally reflect the asymmetry in how we force the MMT equation starting at long wavelengths, our results also provide a unique and detailed nonlinear statistical view into the complex balance of energy transfer across wavenumbers necessary to maintain near statistically steady states in dispersive systems.  

The present work then provides a thorough methodology for analyzing chaotic up to turbulent time series and gives insight into the complexity of stationary cascade formation in multi-wave mixing systems.  Natural next questions for this work are how it performs on other versions MMT model which exhibits breakdown of WWT through the formation of coherent structures \cite{zakharov, RUMPF2005188, PhysRevLett.103.074502}.  Likewise, it would be interesting to study our method in multidimensional WWT models as well as more classically turbulent problems coming from fluid mechanics.  This would also provide a ready space for comparison to the methods in \cite{lozano2, Arranz_Lozano-Durán_2024}.  Ultimately of course, we would also want to see how our approach fares with noisy and incomplete real world measurement.  These are all questions of active research by our group.   

The structure of the paper is as follows.  In Section 2, we present an explanation of transfer entropy, conditional mutual information, and the algorithm we use to compute these quantities in a locally optimal way.  We likewise look at a typical example of its use.  In Section 3, we then present our results on the MMT model.  In Section 4, we provide summary, discussion, and suggest several further directions of research.  

\section{Computing Information Flow in Dynamical Systems}
Given a multidimensional time series, $\left\{{\bf x}_{j}\right\}_{j=0}^{N_{T}}$, with ${\bf x}_{j}\in \mathbb{R}^{N_{d}}$ with vector components denoted as $x_{n,j}$, it is a basic question to determine the extent to which a time series along one dimension {\it causes}, or more broadly {\it informs}, another.  Motivated by the now celebrated Granger causality test, \cf \cite{granger}, in linear time series, \cite{schreiber} introduced the notion of {\it transfer entropy} (TE) to determine the causal relationship between two time series.  The TE from $x_{l,j}$ to $x_{n,j}$, say $T_{x_{l}\rightarrow x_{n}}(j)$ is defined in \cite{schreiber} to be 
\[
T_{x_{l}\rightarrow x_{k}} = H\left(\left.x_{n,j+1}\right|x_{n,j}\right) - H\left(\left.x_{n,j+1}\right|x_{n,j},x_{l,j}\right),
\]
where $H(Y|X)$ is the conditional entropy between two random variables $X$ and $Y$ defined as  
\[
H(Y|X) = \int p(y,x) \log p(y|x) dx dy.
\]
For random variables $X$ and $Y$, we define the information between them $I(X,Y)$ as 
\begin{align*}
I(X,Y) = & H(X) - H(X|Y) \\
= & H(Y) - H(Y|X)\\
= & H(X)+H(Y)-H(X,Y).  
\end{align*}
From this, we see that 
\begin{equation}
T_{x_{l}\rightarrow x_{n}} = I\left(x_{n,j+1}, (x_{n,j},x_{l,j})\right) - I\left(x_{n,j+1}, x_{n,j}\right). 
\label{te_info}
\end{equation}
Given that $I(X,Y)\geq 0$ and $I(X,Y)=0$ if and only if $X$ and $Y$ are independent random variables \cite{calin}, one can show that $T_{x_{l}\rightarrow x_{n}} \geq 0$ and is zero if and only if the time series are independent.  The point then of introducing TE comes from the fact that, for random variables $X$ and $Y$, information is symmetric, i.e. $I(X,Y)=I(Y,X)$.  Thus,  the definition of TE generically satisfies $T_{x_{l}\rightarrow x_{n}} \neq T_{x_{n}\rightarrow x_{l}}$, so that TE allows one to meaningfully track information flow from one variable to another in a multidimensional time series while at the same time allowing one to assess statistical independence between dimensions.      

\subsection{The sIDTxl Algorithm}
Equation \eqref{te_info} shows that TE is equal to the {\it conditional-mutual information} (CMI), say $I(x_{k,j+1},x_{l,j}|x_{k, j})$, between $x_{l}$ and $x_{k}$, so that  
\begin{align*}
T_{x_{l}\rightarrow x_{n}} = & I\left(x_{n,j+1}, (x_{n,j},x_{l,j})\right) - I\left(x_{n,j+1}, x_{n,j}\right) \\
= & I(x_{n,j+1},x_{l,j}|x_{n, j}).
\end{align*}
Recasting TE as CMI has given rise to a host of modifications and improvements on the initial TE concept, see in particular \cite{faes} and \cite{bollt}, which has generated algorithms which can determine networks of interactions between time series that accurately account for confounding variables and non-Markovian influences of past states.   
In particular, in this work, we used the IDTxl library \cite{idtxl} as a template from which to expand.  The backbone of the methods we implement couples the power of non-uniform embeddings of time series \cite{takens, vlachos, faes}, with greedy-algorithm optimization routines which seek out those time series models which provide the most CMI.  The algorithm generates two models.  Fixing a maximum allowable lag $L_{M}$, one model is for {\it sources} in which we find the maximum information flow to $x_{n,j+1}$ from ${\bf x}_{n,\boldsymbol{\ell}^{(n)}_{s}}$, where $\boldsymbol{\ell}^{(n)}_{s}$ represents an optimal choice of some $u$ lags, say $\boldsymbol{\ell}^{(n)}_{s}=(\ell_{1}, \cdots, \ell_{u})$ so that 
\[
{\bf x}^{(S)}_{n,-\lss^{(n)}} = \left\{(x_{n,j-\ell_{1}},\cdots, x_{n,j-\ell_{u}})\right\}_{j=S}^{N_{T}}.
\]
The other lag model that the method generates is for {\it targets} across all complementary dimensions with respect to $n$.  We denote this as $\ltt^{(n)}$ where
\[
\ltt^{(n)} = \cup_{l\neq n} \ltt^{(n,l)}
\]
with, for some subset of $u_{l}$ lags, $\ltt^{(n,l)}=\left(\ell_{1},\cdots,\ell_{u_{l}}\right)$ and 
\[
{\bf x}^{(S)}_{l,-\ltt^{(n,l)}} = \left\{(x_{l,j-\ell_{1}},\cdots, x_{l,j-\ell_{u_{l}}})\right\}_{j=S}^{N_{T}}.
\]
The choice of target lags can vary from target dimension to target dimension, and thus the algorithm is able to find sophisticated non-uniform time embeddings in order to determine information flow within multidimensional time series.  

Each model generation consists of two phases, the first being a BUILD phase, the second being a PRUNE phase.  During BUILD for sources, for dimension $x_{n, j}$, given an existing set of chosen lags say $\lss^{(n)}$ and a set of candidate lags $\boldsymbol{\ell}_{r}$, we then find 
\[
\ell_{\ast} = \text{arg max}_{\ell_{c}\in \boldsymbol{\ell}_{r}} I\left(\left. {\bf x}_{n,L_{c}+1}, {\bf x}^{(L_{c}+1)}_{n,-\ell_{c}} \right|{\bf x}^{(L_{c}+1)}_{n,-\lss^{(n)} } \right), 
\]
with 
\[
{\bf x}_{n,L_{c}+1} = \left\{x_{n,j}\right\}_{j=L_{c}+1}^{N_{T}}, ~ L_{c} = \text{max}\left\{\lss^{(n)} \cup\ell_{c}\right\}.
\]
We then test $\ell_{\ast}$ for statistical significance before expanding the lag model.  After finishing a BUILD phase, in a corresponding PRUNE phase for a source, we compute
\[
\tilde{\ell}_{\ast} = \text{arg min}_{l_{c}\in\lss^{(n)}} I\left(\left. {\bf x}_{n,L_{c}+1}, {\bf x}^{(L_{c}+1)}_{n,-\ell_{c}}  \right| {\bf x}^{(L_{c}+1)}_{n,-\lss^{(n)} \backslash \ell_{c}} \right), 
\]
and then test if $\tilde{\ell}_{\ast}$ is not statistically significant before possibly contracting the lag model.  

Target lag models are then constructed relative to a given source model.  However, given that the algorithm is not guaranteed to generate a globally optimal model, this means that the order in which targets are examined relative to a given source model can have an impact on the result of the model, especially if the information landscape is relatively flat.  To address this issue, we compute and then sort the initial target-to-source CMI values $I^{t,i}_{nl}$ where for $l\neq n$,
\[
I^{t,i}_{nl} = I\left({\bf x}_{n, L_{s}+1}, {\bf x}^{(L_{s}+1)}_{l, -L_{s}} | {\bf x}^{(L_{s}+1)}_{n, -\lss^{(n)}}\right), ~ L_{s} = \text{max}\left\{
\lss^{(n)}\right\}.
\]
Note, we choose to compute $I^{t,i}_{nl}$ with maximal lag separation between target-to-source so as to prioritize first examining those targets with the most significant long-term information couplings to the source.  The algorithm can then potentially replace these longer term couplings with short-term ones, but through our sort we are sure to examine longer term couplings before discarding them. We then iterate over the targets from most to least informative to a given source model.  This prioritizes building target models from the most informative targets to the least, thereby helping to mitigate potentially spurious local optima that the greedy algorithm would otherwise find.  Likewise, it addresses the arbitrary ordering of targets by index and instead allows the algorithm to proceed in a more principled fashion.  

Target lag models are then constructed, though now we include the source model and all previously chosen target models for all priors when computing the CMI.  Thus, during a BUILD phase, after $M$ prior BUILD/PRUNE phases across targets, given source model $\lss^{(n)}$ and prior chosen target-lag models $\ltt^{(n,l_{m})}$, $m=1,\cdots, M$, if the candidate target dimension is currently $l\neq n, l_{1}, \cdots, l_{M}$, with candidate lag model $\ltt^{(n,l)}$, we compute  
\begin{multline*}    
\ell_{\ast} = \text{arg max}_{\ell_{c}\in \boldsymbol{\ell}_{r}} I\left(\left. {\bf x}_{n,L_{c}+1}, {\bf x}^{(L_{c}+1)}_{l,-\ell_{c}} \right|{\bf x}^{(L_{c}+1)}_{n, -\lss^{(n)}}, \right. \\
\left.{\bf x}^{(L_{c}+1)}_{l,-\ltt^{(n,l)}}, {\bf x}^{(L_{c}+1)}_{l_{1},-\ltt^{(n,l_{1})}}, \cdots, {\bf x}^{(L_{c}+1)}_{l_{M},-\ltt^{(n,l_{M})}}  \right), 
\end{multline*}
with $L_{c} = \text{max}\left\{ \lss^{(n)}, \ell_{c}, \ltt^{(n,l)}, \cup_{m=1}^{M} \ltt^{(n,l_{m})} \right\}$.  PRUNE follows in a similar fashion.  For those $\ell_{c}$ which are kept after a PRUNE phase and thus become members of $\ltt^{(n,l)}$, we track the marginal information transfer from target $l$ to source $n$ via the values $I^{m,\ell_{c}}_{nl}$ where 
\begin{multline*}
I^{\ell_{c},m}_{nl} = I\left(\left. {\bf x}_{n,L_{c}+1}, {\bf x}^{(L_{c}+1)}_{l,-\ell_{c}} \right|{\bf x}^{(L_{c}+1)}_{n, -\lss^{(n)}}, {\bf x}^{(L_{c}+1)}_{l,-\ltt^{(n,l)}}, \right.\\ \left. {\bf x}^{(L_{c}+1)}_{l_{1},-\ltt^{(n,l_{1})}}, \cdots, {\bf x}^{(L_{c}+1)}_{l_{M},-\ltt^{(n,l_{M})}}  \right).
\end{multline*}
We likewise find the maximum marginal information transfer $I^{t,m}_{nl}$ via the formula 
\begin{equation}
I^{t,m}_{nl} = \max_{\ell_{c}\in \ltt^{(n,l)}} I^{\ell_{c},m}_{nl}.  
\label{max_marg}
\end{equation}

Before a model is expanded or contracted, respectively whether we are in a BUILD or a PRUNE phase, a proposed choice is tested for statistical significance.  Relative to a testing level $\alpha_{h}$, we determine significance by computing $I(f,S_{m}(c)|b)$ where $f$ is the forward time series that we are trying to model, $c$ is candidate model addition, $b$ is the existing prior model of $f$, and $S_{m}(c)$ is the $m^{th}$ random shuffle in time of $c$.  Over $N_{sh}$ shuffles, we find
\[
I_{sh} = \max_{1\leq m \leq N_{sh}} I(f,S_{m}(c)|b).   
\]
We then generate $N_{sig} = 1/\alpha_{h}+1$ values $I^{(l)}_{sh}$ and define, for a BUILD phase, the value $p_{h}$ to be 
\[
p_{h} = \frac{1}{N_{sig}}\sum_{l=1}^{N_{sig}}\mathcal{I}\left( I^{(l)}_{sh} - I(f,c|b)\right)
\]
and for a PRUNE phase
\[
p_{h} = \frac{1}{N_{sig}}\sum_{l=1}^{N_{sig}}\mathcal{I}\left(I(f,c|b) - I^{(l)}_{sh} \right)
\]
where 
\[
\mathcal{I}(x) = \left\{
\begin{array}{rl}
1 & x \geq 0 \\
0 & x < 0.
\end{array}
\right.
\]
We accept a potential BUILD/PRUNE change if $p_{h}<\alpha_{h}$.  Note, the use of a shuffled candidate $S_{m}(c)$ should ensure that $I(f,S_{m}(c)|b) \ll 1$ and thus $I_{sh}\ll 1$.  The choice of $N_{sig}=1/\alpha_{h}+1$ then ensures that $p_{h} > \alpha_{h}$ if a value $I(f,c|b)$ is either too small or too large relative to at least two values of $I_{sh}^{(l)}$.  Throughout our simulations, we choose $N_{sh}$ (the number of shuffles) to be 
\[
N_{sh} = \text{max}\left\{10, \mathcal{N}(b)\right\}
\]
where $\mathcal{N}(b)$ is the number of lagged target and source model terms at the present stage of the sIDTxl algorithm.  

To compute the CMI, we use nearest-neighbor estimators developed in \cite{grassberger}, which we label the KSG estimator.  Here, one must choose the number of nearest neighbors, say $k_{n}$, to be used in the estimators.  By following the practice laid out in \cite{grassberger} of scaling all time series to have zero mean and unit variance, the overall results are not especially sensitive to the choice of $k_{n}$.  The greater dilemma then is the sample size of the data used, \cf \cite{bollt3}.  Through trial and error we generally found $k_{n}=3$ to consistently produce the best results.  Arguably, the impact of the choice of $k_{n}$ is greatest when we determine statistical significance in the sIDTxl algorithm.  All of these processes are summarized in Algorithm \ref{idtxl}; for additional details see \cite{idtxl2}.  

\begin{algorithm}[H]
\caption{sIDTxl (Sorted IDTxl) Algorithm}
\label{idtxl}
\begin{algorithmic}[1]
	\For{Dimension n}
	\Procedure{Generate Source Model for $x_{n,j}$}{}	
        \State INITIALIZE: $\lss^{(n)} = \left\{\varnothing\right\}$, $\boldsymbol{\ell}_{r}=\left\{1\cdots L_{M}\right\}$.        
	\Procedure{Build}{}
		\While{$\boldsymbol{\ell}_{r}\neq \left\{\varnothing\right\}$}
        \State Given $\lss^{(n)} =\left\{\ell_{1} \cdots \ell_{c}\right\}$ 
        \State Given $\boldsymbol{\ell}_{r}=\left\{ 1,\cdots,L_{M}\right\}\backslash \lss $        
        \State $
        \ell_{\ast} \leftarrow \text{arg max}_{\ell_{c+1}\in \boldsymbol{\ell}_{r}} \tilde{I}$
        \State with $\tilde{I} = I\left(\left. {\bf x}_{n,L_{c}+1}, {\bf x}^{(L_{c}+1)}_{n,-\ell_{c+1}} \right|{\bf x}^{(L_{c}+1)}_{n, -\lss^{(n)} } \right)$             	
        
     	\If{$\ell_{\ast}$ is statistically significant}
     	\State $\lss^{(n)} \leftarrow \lss^{(n)} \cup \left\{\ell_{\ast}\right\}$
     	\EndIf
     	\EndWhile     	
    \EndProcedure
    \Procedure{Prune}{INITIALIZE: $S \equiv \text{True}$}
        \While{$S$}
        	\State $\tilde{\ell}_{\ast}\leftarrow \text{arg min}_{l_{c}\in\lss^{(n)}} \tilde{I}$

            \State with $\tilde{I}=I\left(\left. {\bf x}_{n,L_{c}+1}, {\bf x}^{(L_{c}+1)}_{n,-\ell_{c}}  \right| {\bf x}^{(L_{c}+1)}_{n,-\lss^{(n)} \backslash \ell_{c}} \right)$
            
        \If{$\tilde{\ell}$ is statistically insignificant}
        \State $\lss^{(n)} \leftarrow \lss^{(n)} \backslash \left\{\tilde{\ell}_{\ast}\right\}$
        \Else
        \State $S\equiv \text{False}$
        \EndIf
        \EndWhile
    \State RETURNS: $\lss^{(n)}$
    \EndProcedure    
    \EndProcedure        
    \Procedure{Generate Target Model for $x_{n,j}$}{}
            \State SORT Targets: for $l \neq n$, compute and sort $I^{t,i}_{nl}$.  
    	\For{ $l\neq n$ following the sort of $I^{t,i}_{nl}$ }
    		\State INITIALIZE: $\ltt^{(n,l)} = \left\{\varnothing\right\}$, $\boldsymbol{\ell}_{r}=\left\{1\cdots L_{M}\right\}$.        
    		\Procedure{Build}{}
    		\State Build $\ltt^{(n,l)}$ from $x_{l,j}$ on $\lss$ and prior $\ltt^{(n,l_{m})}$.
    		\EndProcedure
    		\Procedure{Prune}{}
    		\State Prune $\ltt^{(n,l)}$ on $\lss^{(n)}$ and prior $\ltt^{(n,l_{m})}$.
    		\State RETURNS: $\ltt^{(n,l)}$
    		\EndProcedure
    	\EndFor
    \EndProcedure
    \State RETURNS: $\lss^{(n)}$, $\cup_{l\neq n}\ltt^{(n,l)}$
    \EndFor    
\end{algorithmic}
\end{algorithm}

In the remainder of the paper, we report the output from the sIDTxl method as: 
\begin{itemize}
\item The initial target-to-source CMI values $I^{t,i}_{nl}$, which provides a comprehensive view of the information landscape between targets to respective sources prior to target selection and model development.
\item The significant maximum marginal values $I^{t,m}_{nl}$ (see Equation \eqref{max_marg}), which shows which targets are selected relative to a given source, and what the marginal additional information is provided by a chosen target and corresponding lag.
\item The final $I^{t,f}_{nl}$ where 
\[
I^{t,f}_{nl} = I\left({\bf x}_{n, L_{s}^{(n)}+1}, {\bf x}^{(L_{s}^{(n)}+1}_{l, -\ltt^{(n)}} | {\bf x}^{(L_{s}^{(n)}+1}_{n, -\lss^{(n)}}\right), ~ l\neq n, 
\]
\[
L^{(n)}_{s} = \text{max}\left\{
\lss^{(n)}, \ltt^{(n)}\right\},
\]
and $\ltt^{(n)}$ is the computed lag model from the targets $\left\{x_{l, j}\right\}_{j=1}^{N_{T}}$, $l\neq n$.  Thus, for chosen targets, $I^{t,f}_{nl}$ is the final target-to-source information after computing the optimal target and source lag models.  
\item The maximum lag for a source or target model, thereby illustrating the various time scales computed by sIDTxl.  
\end{itemize}
\subsection{Test Cases and Algorithm Analysis}
\begin{figure}
\centering
\begin{tabular}{cc}
\includegraphics[width=.22\textwidth]{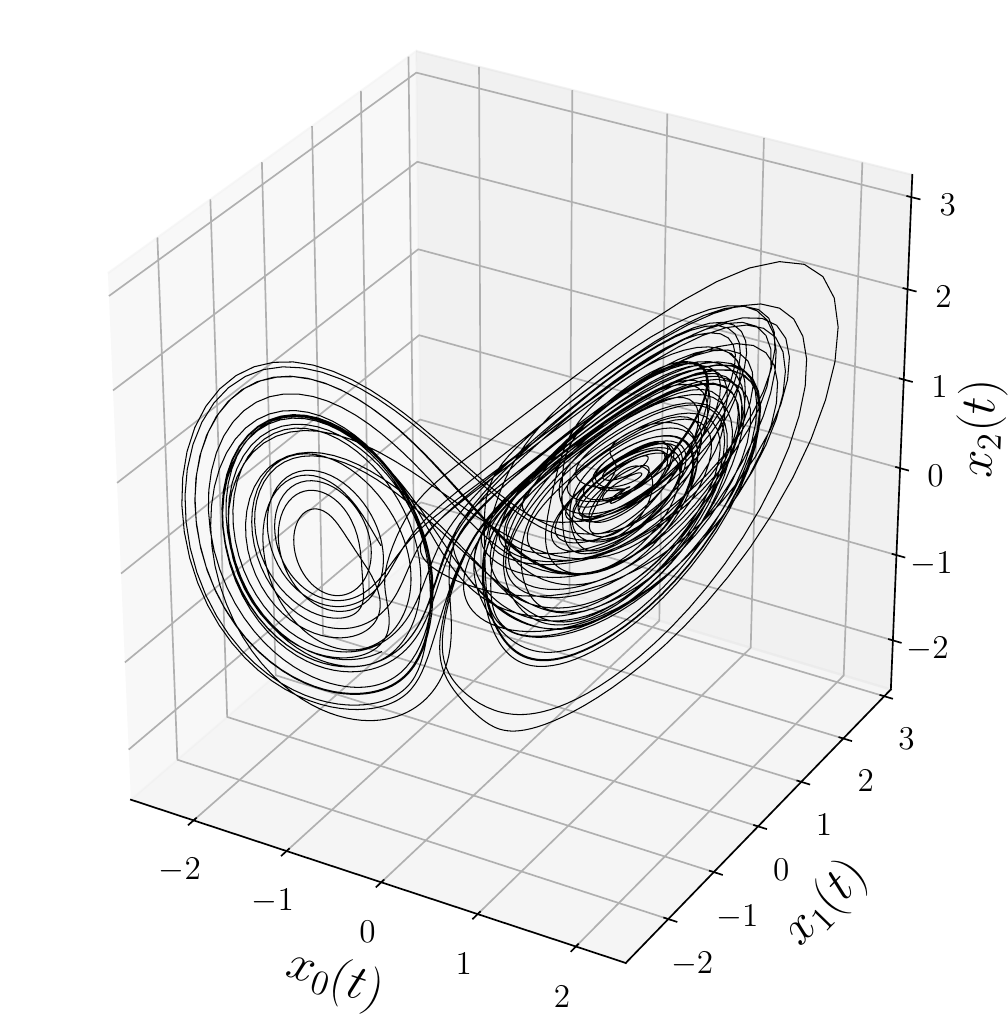} & \includegraphics[width=.22\textwidth]{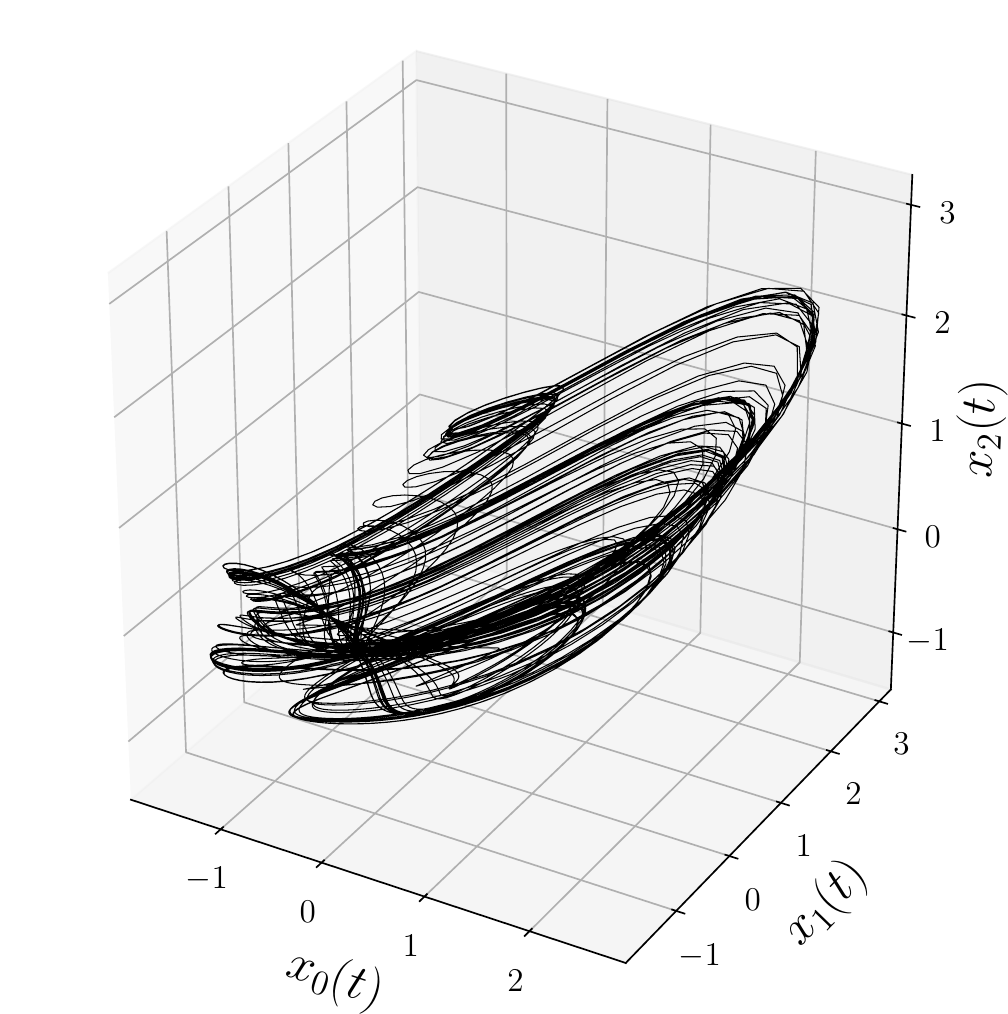}\\
(a) $C$ = 1. & (b) $C$=10.\\
\includegraphics[width=.22\textwidth]{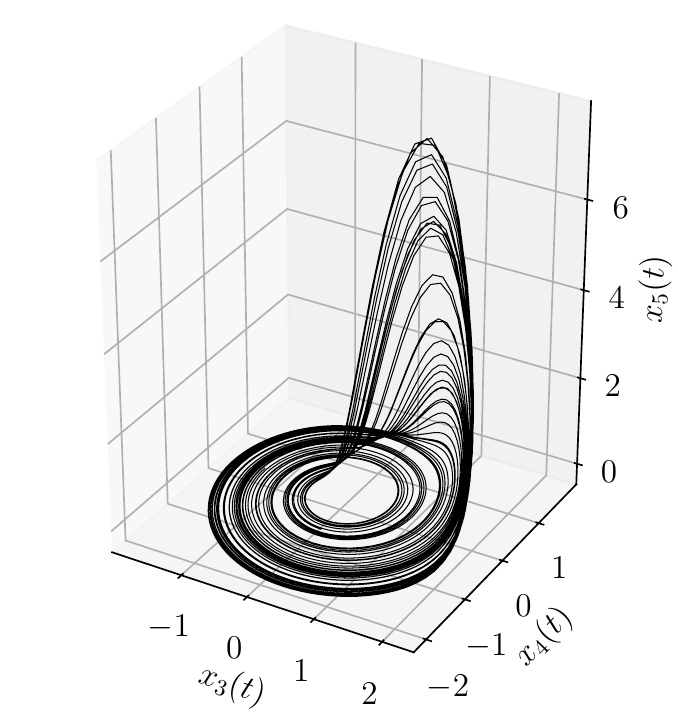} & \\
(c) & 
\end{tabular}
\caption{Plot of Lorenz--R\"{o}ssler system.  Top Figures: Lorenz dynamics for $C=1$ (a) and $C=10$ (b).  Bottom Figure: R\"{o}ssler dynamics (c).}
\label{fig:coupled_sys_dynamics}
\end{figure}

To explore the use of sIDTxl and its related issues, we study a common problem from the affiliated literature, which is the coupled Lorenz--R\"{o}ssler system of the form 
\begin{align*}
\dot{x}_{0} = & \sigma(x_{1}-x_{0})\\
\dot{x}_{1} = & x_{0}(\rho-x_{2}) - x_{1} + Cx_{4}^{2} \\
\dot{x}_{2} = & x_{0}x_{1} - \beta x_{2}\\
\dot{x}_{3} = & -6(x_{4}+x_{5}) \\
\dot{x}_{4} = & 6(x_{3} + \alpha x_{4})\\
\dot{x}_{5} = & 6(\gamma + x_{5}(x_{3}-\delta))\\
\end{align*}
Here we let $\sigma=10$, $\rho=28$, $\beta=8/3$, $\alpha=.2$, $\gamma=.2$, $\delta=5.7$.  $C$ can be varied so as to enhance the driving effect of the R\"{o}ssler system on the Lorenz system, though the effect of this can be surprising, especially when looked at over the whole network; see Figure \ref{fig:coupled_sys_dynamics} for details.
Throughout our tests, we use trajectories found via a 4th-order Runge--Kutta scheme using a time step of $\delta t = .01$ run out to a total time of $150$ units of non-dimensional time.  The first $120$ units of time are ignored so as to remove any transient phenomena from our data set.  For $C=1$, using 100 randomly chosen initial conditions, we find the average maximum Lyapunov exponent, say $\lambda_{M}$, to be $\lambda_{M}\approx.54$.  Likewise, for $C=10$, we find $\lambda_{M}=.42$.  Note, the Lyapunov exponents are computed using the methods in \cite{dieci}.  Thus, the length of data we use for the sIDTxl algorithm consists of a significant number of time intervals over which chaotic dynamics can occur.  To make computations run in more reasonable amounts of time, we subsample our time series so that it has a time step of $\delta t= .05$.    

\begin{figure}
\centering
\begin{tabular}{c}
\includegraphics[width=.49\textwidth]{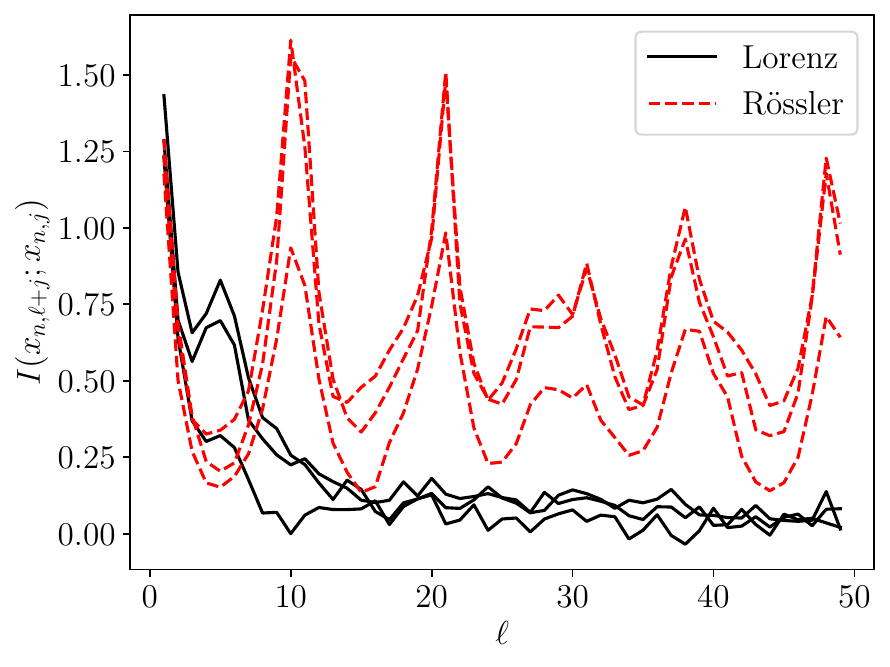} \\
(a) $C=1$, $\lambda_{M}=.54$ \\
\includegraphics[width=.49\textwidth]{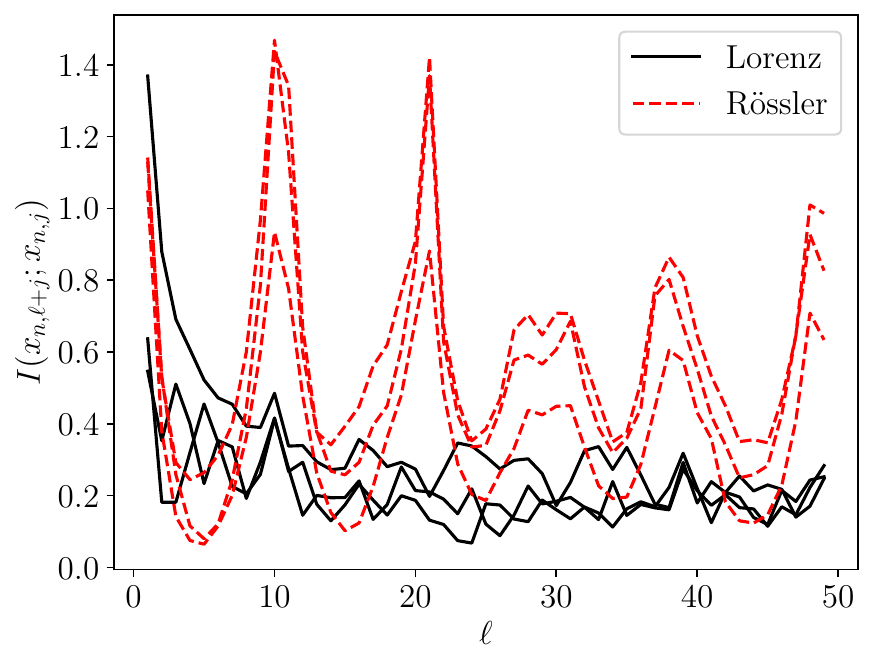}\\
(b) $C=10$, $\lambda_{M}=.42$ 
\end{tabular}
\caption{Information across lags $\ell$, $I\left(x_{n, \ell+j}, x_{n}\right)$, for the Lorenz--R\"{o}ssler system for $C=1$ (a) and $C=10$ (b) with the maximum lag corresponding to the reciprocal of the average maximum Lyapunov exponent $\lambda_{M}$.}
\label{fig:lr_lag_test}
\end{figure}

Before proceeding further, we establish a principled means of determining a maximal lag $L_{M}$.  To do so, we compute the quantities $I\left(x_{n, \ell + j}, x_{n}\right)$ for $\ell\geq 1$ up to $1/\lambda_{M}$; the results for $C=1$ and $C=10$ are shown in Figure \ref{fig:lr_lag_test}.  As can be seen, the near-periodicity of the R\"{o}ssler system manifests in the spikes in information starting at $\ell=10$ and then periodically continuing.  Likewise, the stronger coupling at $C=10$ causes the Lorenz system to have information oscillations across lags reflecting the strength of the driving and concomitant entrainment.  For $C=1$ and $C=10$, we chose $L_{M}=3$ since it is the first local minimum thereby capturing the most informative lags along the Lorenz coordinate but before oscillations in the information cause new local peaks.  On this point, larger choices of $L_{M}$ tended to miss any impacts from targets due to their effects being washed out by the oscillating self-information contained in the target model.  This approach echoes the methods proposed in \cite{swinney} which used information theory to find optimal lags for computing Takens' embeddings \cite{takens}.  In this line of thinking, one sees a tension of needing enough lags so that appropriate time scales of inter-dimensional couplings are represented while not having so many choices of lags that one can fully explain the dynamics through source lag embeddings alone.  

\begin{figure}
\centering
\begin{tabular}{cc}
\includegraphics[width=.24\textwidth]{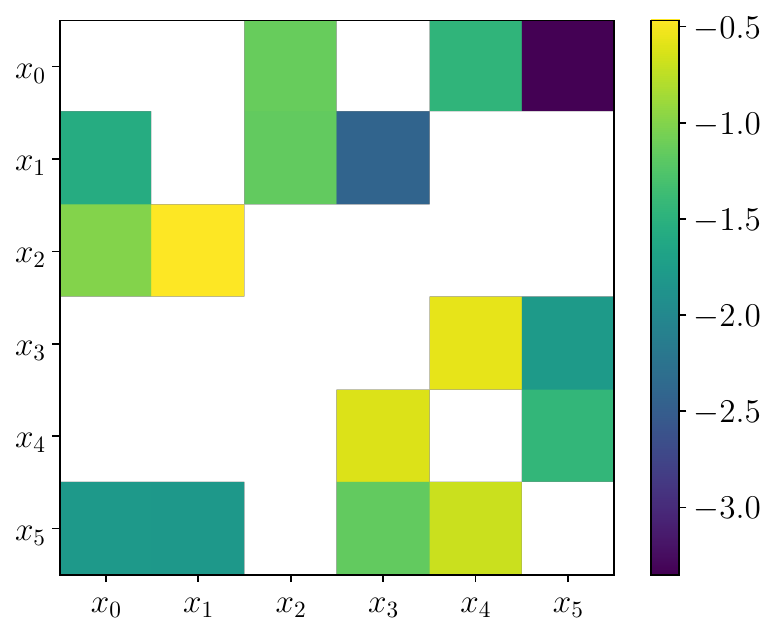} & 
\includegraphics[width=.24\textwidth]{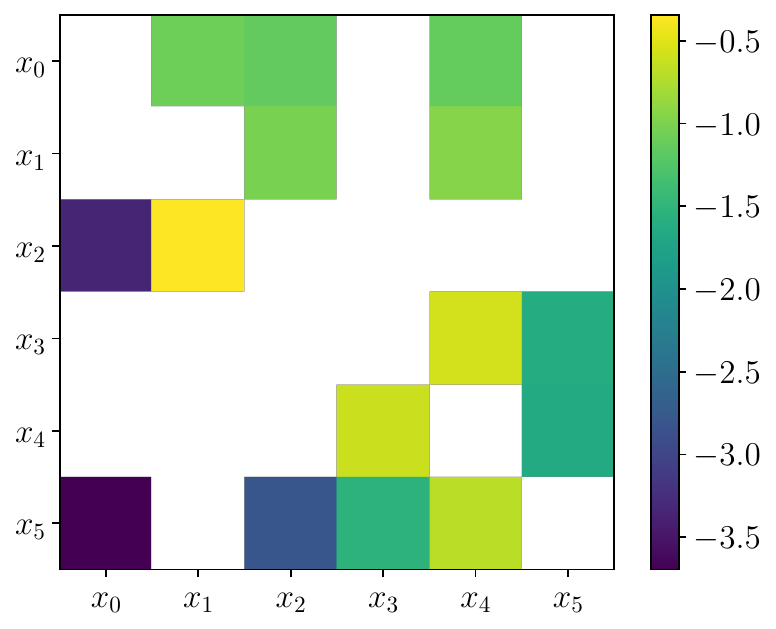}\\
(a) $I^{t,i}$ & (b) $I^{t,m}$\\ 
\includegraphics[width=.24\textwidth]{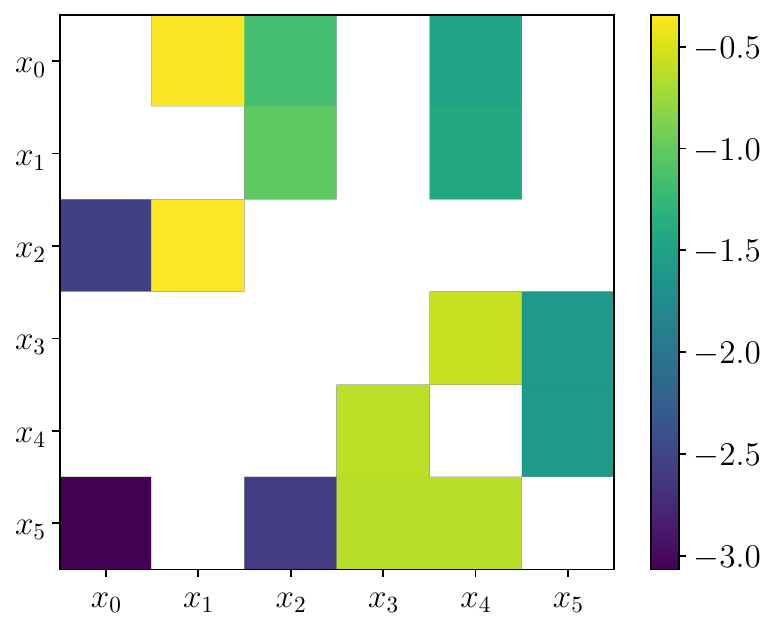} & \includegraphics[width=.24\textwidth]{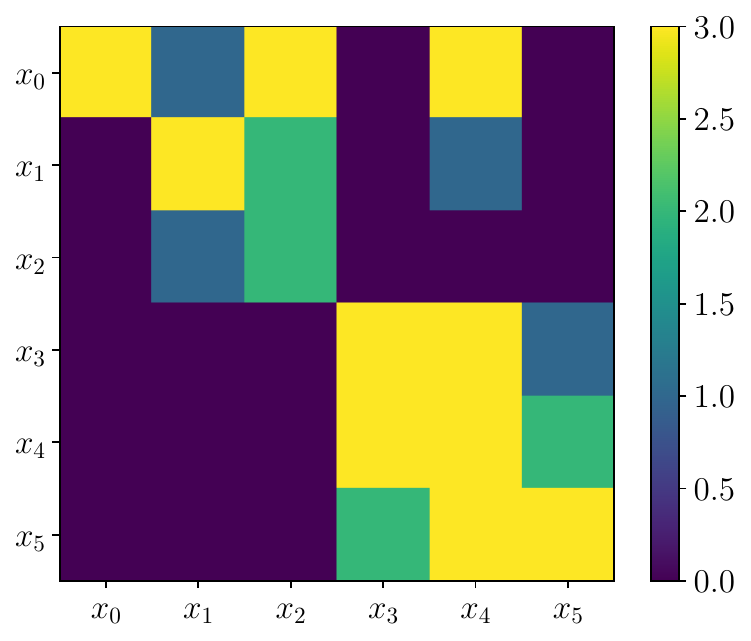}\\
(c) $I^{t,f}$ & (d) Max Lags\\
\includegraphics[width=.24\textwidth]{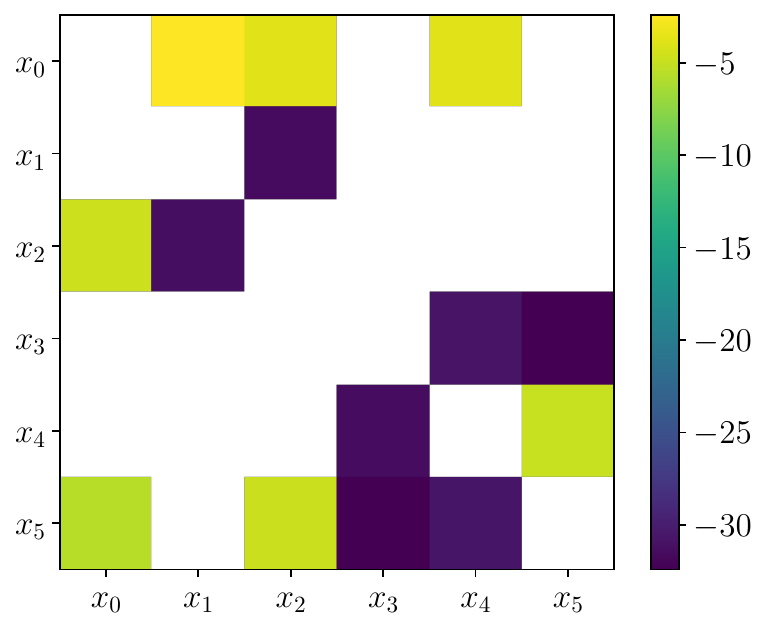} & 
\includegraphics[width=.24\textwidth]{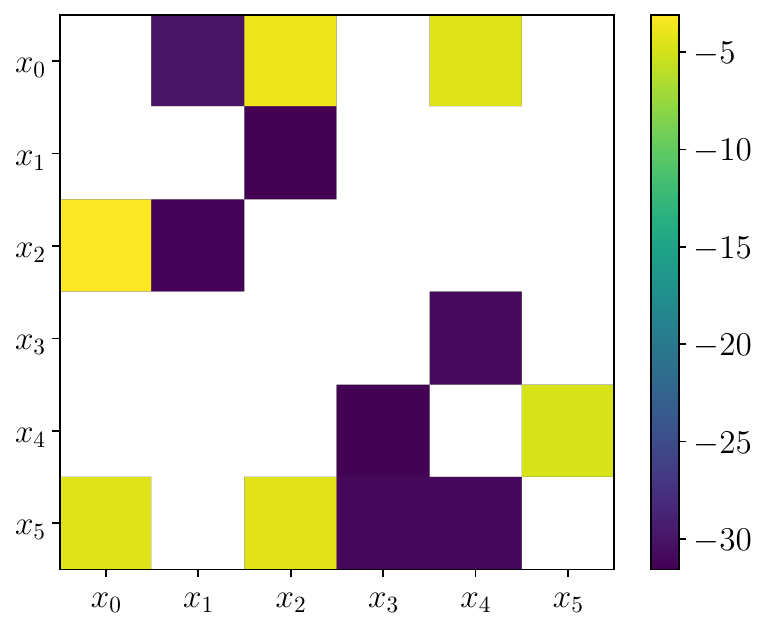}\\
(e) $\sigma(I^{t,m})$ & (f) $\sigma(I^{t,f})$
\end{tabular}
\caption{ The average values over $N_{tr}=100$ trials of (a) $I^{t,i}$, (b) $I^{t,m}$,  (c) $I^{t,f}$, and (d) the maximum lags for the Lorenz--R\"{o}ssler system for $C=1$.  The standard deviations around the averages of $I^{t,m}$ and $I^{t,f}$ are seen in (e) and (f).  Figures (a)-(c), (e), and (f) are plotted on a log-scale.  The hypothesis-testing threshold is $\alpha_{h}=.05$, the maximum allowed lag is $L_{M}=3$, and the number of nearest neighbors is $k_{n}=3$.  Note, coupling from target-to-source is read left-to-right.}
\label{fig:coupled_sys_C_1}
\end{figure}

With the choice of $L_{M}=3$ now set, and letting $\alpha_{h}=.05$, we run sIDTxl for $N_{tr}=100$ trials and compute the averages of each relevant quantity.  For coupling strength $C=1$, we get the results in Figure \ref{fig:coupled_sys_C_1}.  We show the average $I^{t,i}$ in Figure \ref{fig:coupled_sys_C_1}(a); average $I^{t,m}$ in Figure \ref{fig:coupled_sys_C_1}(b); average $I^{t,f}$ in Figure \ref{fig:coupled_sys_C_1}(c), and the average maximum lag determined by the sIDTxl algorithm in Figure \ref{fig:coupled_sys_C_1}(d).  The variances around the averages in Figures \ref{fig:coupled_sys_C_1}(b) and (c) are plotted in Figures \ref{fig:coupled_sys_C_1} (e) and (f).  From Figure \ref{fig:coupled_sys_C_1}(a) alone then, we see that information largely clusters along each isolated dynamical system with relatively weak couplings across the two systems.  However, in Figures \ref{fig:coupled_sys_C_1}(b) and (c) we see that the method is able to isolate that $x_{4}$ is the driving dimension, though it couples to both $x_{0}$ and $x_{1}$.  It also finds weak couplings from $x_{0}$ and $x_{2}$ to $x_{5}$.  Looking at the variances in Figures \ref{fig:coupled_sys_C_1} (e) and (f) though, we see the false positive from $x_{4}$ to $x_{0}$ is of markedly higher variance than the true coupling from $x_{4}$ to $x_{0}$.  Thus we have good reason to ignore the incorrect coupling.  A similar argument holds for the false positive couplings from $x_{0}$ and $x_{2}$ to $x_{5}$.  We also note that $x_{5}$ represents a near singular perturbation in the R\"{o}ssler system, so that the sIDTxl method struggles in this case is not so surprising.  

Nevertheless, we see that the sIDTxl method is able to see through what would be otherwise confounding variables to detect couplings across the two dynamical systems.  Refering to the maximal lag choices in Figure \ref{fig:coupled_sys_C_1}(d), we also see in $I^{t,f}$ that by allowing targets to use smaller lags that some couplings are found to be more informatively rich while other couplings computed at larger lags in $I^{t,i}$ are discarded.  Likewise, we see that the method finds a variety of largest lag values showing that different time scales influence information flow.  In particular, the $x_{4}$ to $x_{1}$ coupling happens essentially over one time step without recourse to longer ones.  In contrast, source-to-source couplings are more informative when allowing for the longest available lags, though in some ways this feature is problematic since letting $L_{M}$ become too large allows the sIDTxl method to ignore targets in favor of longer source models.    

\begin{figure}
\centering
\begin{tabular}{cc}
\includegraphics[width=.24\textwidth]{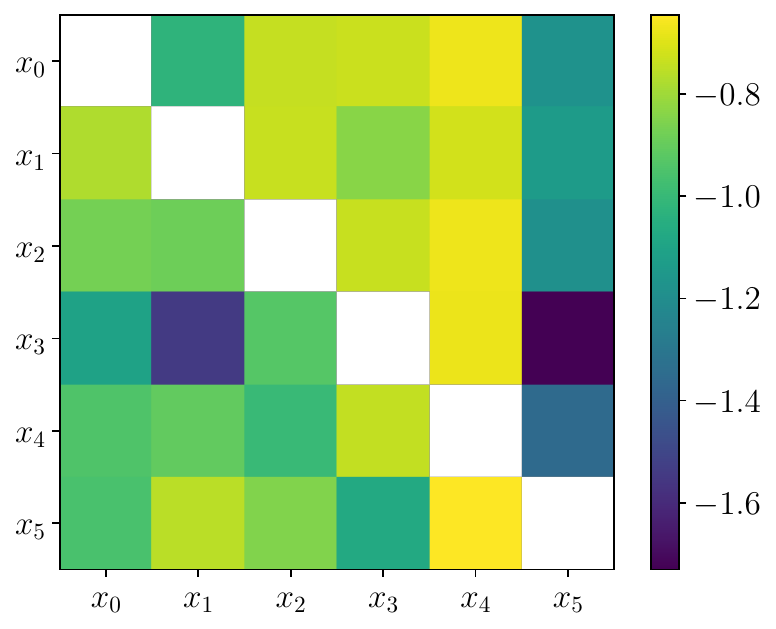} & 
\includegraphics[width=.24\textwidth]{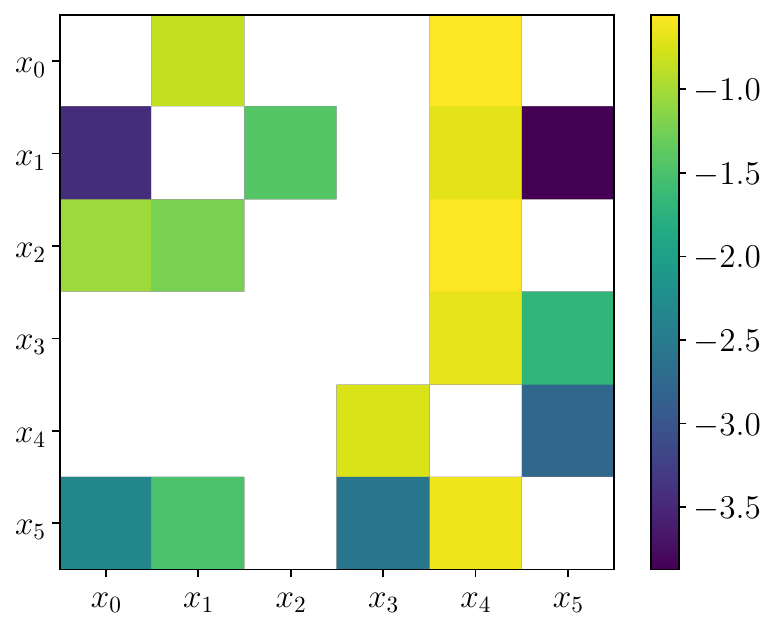}\\
(a) $I^{t,i}$ & (b) $I^{t, m}$\\ 
\includegraphics[width=.24\textwidth]{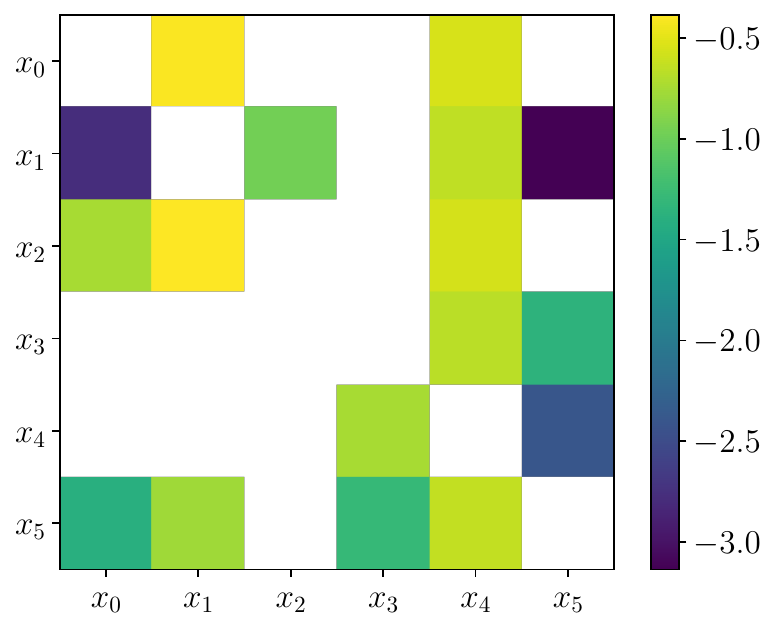} & \includegraphics[width=.24\textwidth]{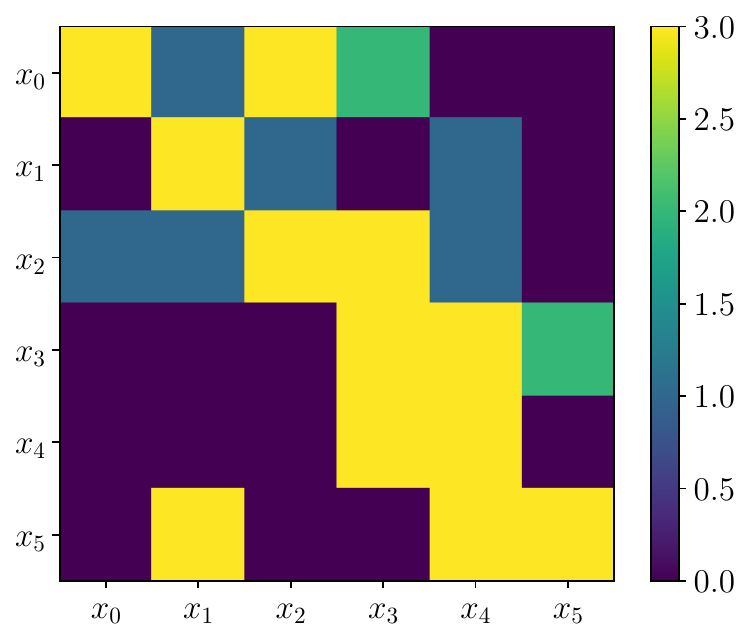}\\
(c) $I^{t,f}$ & (d) Max Lags\\
\includegraphics[width=.24\textwidth]{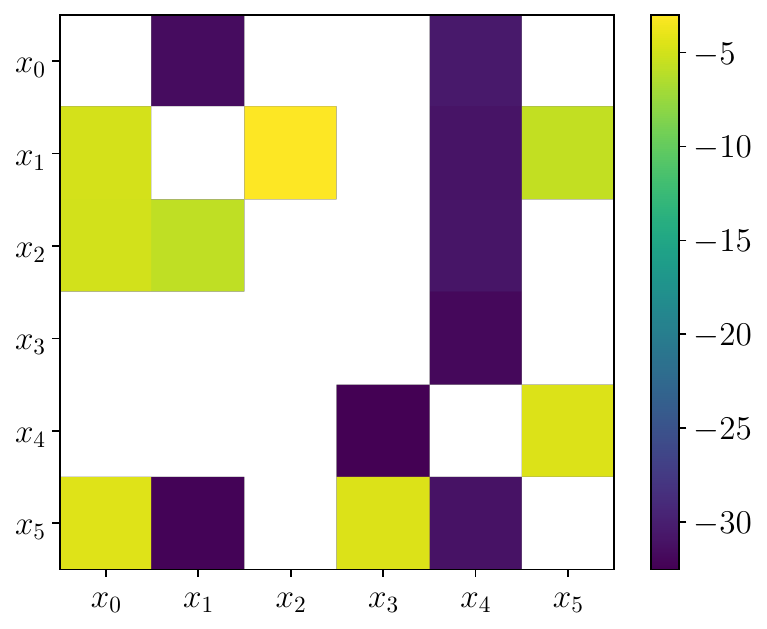} & 
\includegraphics[width=.24\textwidth]{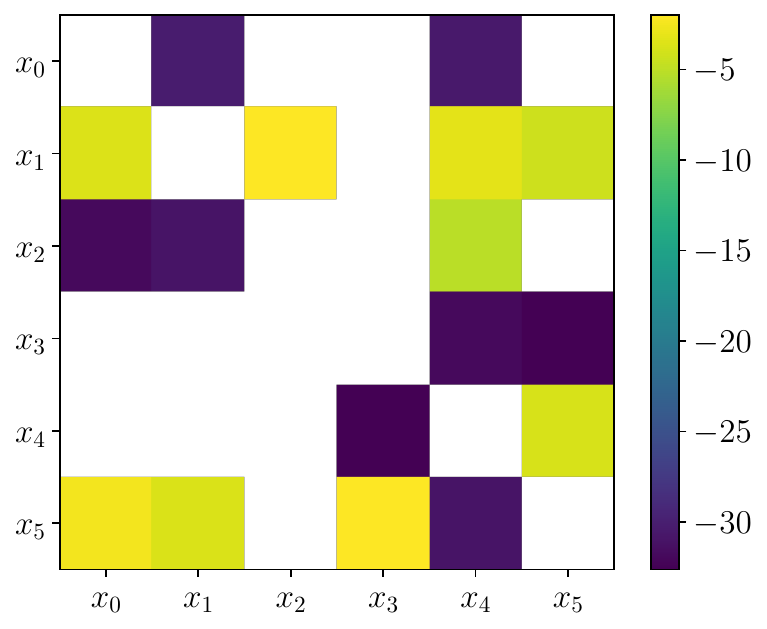}\\
(e) $\sigma(I^{t,m})$ & (f) (e) $\sigma(I^{t,f})$
\end{tabular}
\caption{ The average values over $N_{tr}=100$ trials of (a) $I^{t,i}$, (b) $I^{t,m}$,  (c) $I^{t,f}$, and (d) the maximum lags for the Lorenz--R\"{o}ssler system for $C=10$.  The standard deviations around the averages of $I^{t,m}$ and $I^{t,f}$ are seen in (e) and (f).  Figures (a)-(c), (e), and (f) are plotted on a log-scale.  The hypothesis-testing threshold is $\alpha_{h}=.05$, the maximum allowed lag is $L_{M}=3$, and the number of nearest neighbors is $k_{n}=3$.  Note, coupling from target-to-source is read left-to-right.}
\label{fig:coupled_sys_C_10}
\end{figure}

We also examine the case of $C=10$ with results presented in Figures \ref{fig:coupled_sys_C_10} (a)-(f).  Immediately, we see in Figure \ref{fig:coupled_sys_C_10} (a) a more complicated information landscape with many more potentially significant couplings which coexist on a more uniform information background.  Correspondingly, we see in Figure \ref{fig:coupled_sys_C_10} (b) that while $x_{4}$ is seen as the most significant driver of the Lorenz system by several orders of magnitude, this coupling is washed out across all of the Lorenz dimensions.  Thus, while the R\"{o}ssler system is still otherwise identified correctly despite the presence of a relatively weaker coupling from $x_{1}$ to $x_{5}$, sIDTxl's decision making with regards to the Lorenz coordinates is muddier.  This reflects the strong distortions of trajectories one sees in Figure \ref{fig:coupled_sys_dynamics} (b), which propagates as strong information couplings across most of the R\"{o}ssler coordinates relative to the Lorenz coordinates.  Thus, we see a limitation in the sIDTxl method relative to very strong couplings across systems which can manifest in a relatively large number of false positives.  Interestingly, these errors come in large part by exploiting the largest possible lag choices as seen in Figure \ref{fig:coupled_sys_C_10} (d).  

%%%%%%%%%%%%%%%%%%%%%%%%%%%%%%%%%%%%%%%%%%%%%%%%%%%%%%%%%%%%%%%%%%%%%%%%%%%%%%%%%%%%%%%%%%%%%%%%
%%%%%%%%%%%%%%%%%%%%%%%%%%%%%%%%%%%%%%%%%%%%%%%%%%%%%%%%%%%%%%%%%%%%%%%%%%%%%%%%%%%%%%%%%%%%%%%%
\section{Information Transfer in a Weakly Turbulent System}
%%%%%%%%%%%%%%%%%%%%%%%%%%%%%%%%%%%%%%%%%%%%%%%%%%%%%%%%%%%%%%%%%%%%%%%%%%%%%%%%%%%%%%%%%%%%%%%%
%%%%%%%%%%%%%%%%%%%%%%%%%%%%%%%%%%%%%%%%%%%%%%%%%%%%%%%%%%%%%%%%%%%%%%%%%%%%%%%%%%%%%%%%%%%%%%%%
We now explore using the sIDTxl library on data coming from the MMT model \cite{majda1}.  The particular MMT model we study is of the form 
\begin{multline*}
i\partial_{t}\psi = \left|\partial_{x}\right|^{1/2}\psi - \left|\psi\right|^{2}\psi \\ 
+ i\epsilon^{2}\left(f-\left(\frac{\left|\partial_{x}\right|}{k^{d}_{+}}\right)^{d_{+}} -\left(\frac{k^{d}_{-}}{\left|\partial_{x}\right|}\right)^{d_{-}} \right)\psi,
\end{multline*}
where the forcing $f$ is defined so that 
\[
\widehat{f\psi}(k,t) = \left(\sum_{\tilde{k}=k_{-}^{f}}^{k_{+}^{f}}\hat{\delta}(k-\tilde{k})\right)\hat{\psi}(k,t),
\]
where $\hat{\delta}(k) = 1$ for $k=0$ and is zero otherwise.  The range of forcing wavenumbers between $k^{f}_{-}$ and $k^{f}_{+}$ define the {\it forcing} regime.  Likewise, we damp long waves for $|k|<k^{d}_{-}$ and short waves for $|k|>k^{d}_{+}$.  Those wavenumbers that are sufficiently greater than $k^{f}_{+}$ but smaller than $k^{d}_{+}$ define the {\it inertial range} denoted by $k^{i}_{-}<k<k^{i}_{+}$.

Our interest then in this model comes from the fact that it is a {\it weakly turbulent} system, which means that it generates spatio-temporally chaotic dynamics which, in a properly identified inertial range, can be described by a statistically steady profile.  Quantitatively, by defining \[
\tilde{n}(k,t) = \left<\left|\hat{\psi}(k,t)\right|^{2} \right>, 
\]
one can show \cite{zakharov} in the long time limit that $\tilde{n}(k,t) \rightarrow \tilde{n}_{eq}(k)$ where $\tilde{n}_{eq}(k) = C|k|^{-1}$.  Within this equilibrium distribution, we should anticipate both inverse and forward cascades by looking at the {\it particle number} $\mathcal{N}$ and {\it energy} $\mathcal{E}$, given respectively by 
\[
\mathcal{N} = \sum_{k} \tilde{n}(k,t), ~ \mathcal{E} = \frac{1}{2}\int \left|\left|\partial_{x}\right|^{1/4}\psi\right|^{2}dx - \frac{1}{4} \int \left|\psi\right|^{4}dx.
\] 
Note, $\mathcal{E}$ is the Hamiltonian for the unforced MMT equation.  For a solution $\psi = \mathcal{O}(\epsilon)$, using Parseval's equality, $\mathcal{E}$ is approximately given by
\[
\mathcal{E} \approx \pi \sum_{k} |k|^{1/2}\tilde{n}(k,t).
\]
Following the argument in \cite{majda1} then, we expect to see `forward' cascades of $\mathcal{E}$ and `inverse' cascades of $\mathcal{N}$ if we force at low wavenumbers.   

However, as explored in \cite{RUMPF2005188, PhysRevLett.103.074502, PhysRevE.95.062225, lvov, hrabski, Simonis_Hrabski_Pan_2024} the process by which statistically stationary conditions is achieved is intricate.  One can see this by ignoring forcing and damping, which is appropriate within the inertial regime, and then passing to a Fourier representation of the MMT model written as 
\begin{multline*}
i\partial_{t}\hat{\psi}_{k}(t) = |k|^{1/2}\hat{\psi}_{k}(t) \\
- \sum_{k_{1},k_{2},k_{3}}\hat{\psi}_{1}(t)\hat{\psi}_{2}(t)\hat{\psi}_{3}^{\ast}(t)\delta\left({\bf k}_{12,3k}\right),
\end{multline*}
where $\hat{\psi}_{n}(t) = \hat{\psi}(k_{n},t)$, $\hat{\psi}_{k}(t) = \hat{\psi}(k,t)$, and ${\bf k}_{12,3k} = k_{1}+k_{2}-k_{3}-k$.  From this, one can readily show that 
\begin{equation}
\partial_{t}\left|\hat{\psi}_{k}(t)\right|^{2} = -2\sum_{k_{1},k_{2},k_{3}}\text{Im}\left(\hat{\psi}_{1}\hat{\psi}_{2}\hat{\psi}_{3}^{\ast}\hat{\psi}_{k}^{\ast}\right)\delta({\bf k}_{12,3k}),
\label{eq:interaction}
\end{equation}
Defining $\omega_{n} = |k_{n}|^{1/2}$ and using the substitution $\hat{\phi}_{n}(t) = \hat{\psi}_{n}(t)e^{i\omega_{n}t} $, we in turn find that 
\[
\partial_{t}\left|\hat{\psi}_{k}(t)\right|^{2} = -2\sum_{k_{1},k_{2},k_{3}}\text{Im}\left(\hat{\phi}_{1}\hat{\phi}_{2}\hat{\phi}_{3}^{\ast}\hat{\phi}_{k}^{\ast}e^{-i\boldsymbol{\omega}_{12,3k}t}\right)\delta({\bf k}_{12,3k}),
\]
with $\boldsymbol{\omega}_{12,3k} = \omega_{1}+\omega_{2}-\omega_{3}-\omega(k)$, and where we have used the fact that $|\hat{\psi}_{k}(t)| = |\hat{\phi}_{k}(t)|$.  
Thus, in the long time limit, a stationary phase argument shows us that those wavenumbers that lead to, or nearly to, {\it 4-wave mixing}, i.e.
\[
k_{1}+k_{2}-k_{3}-k=0, ~ \omega_{1}+\omega_{2}-\omega_{3}-\omega(k)=0,
\]
drive the process of convergence and maintenance of a statistically steady state.  Therefore, we can have significant multiscale energy transfer across otherwise widely separated scales.  This greatly complicates the question of tracking information flow, and having some quantitative sketch of this process is of interest.  

\begin{figure}
\centering
\includegraphics[width=.45\textwidth]{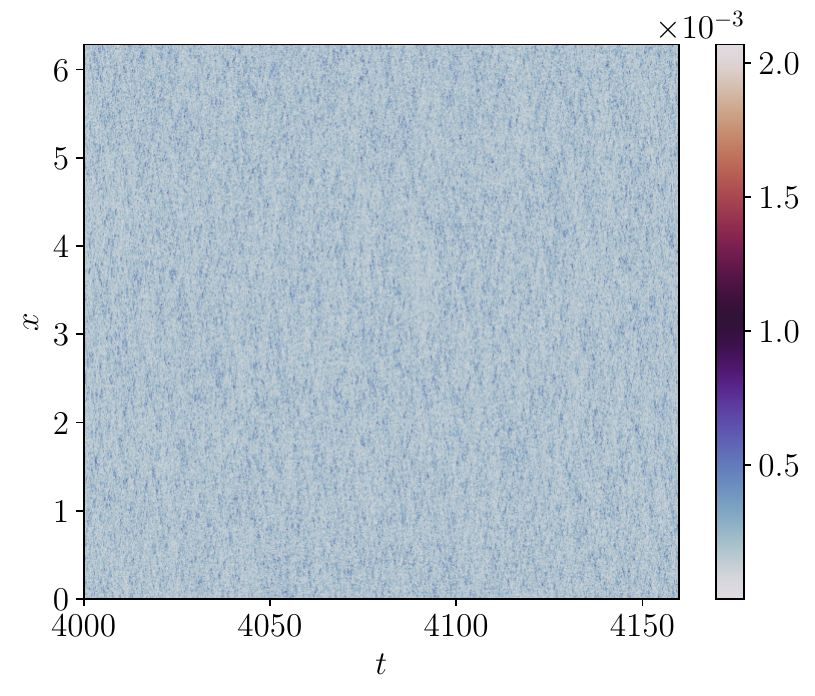}
\caption{Plot of $|\psi(x,t)|$ for $\frac{2k^{i}_{+}}{\epsilon^{2}} < t < \frac{2k^{i}_{+}}{\epsilon^{2}} + 160$.}
\label{fig:mmtdynamics}
\end{figure}

\begin{figure}
\centering
\includegraphics[width=.4\textwidth]{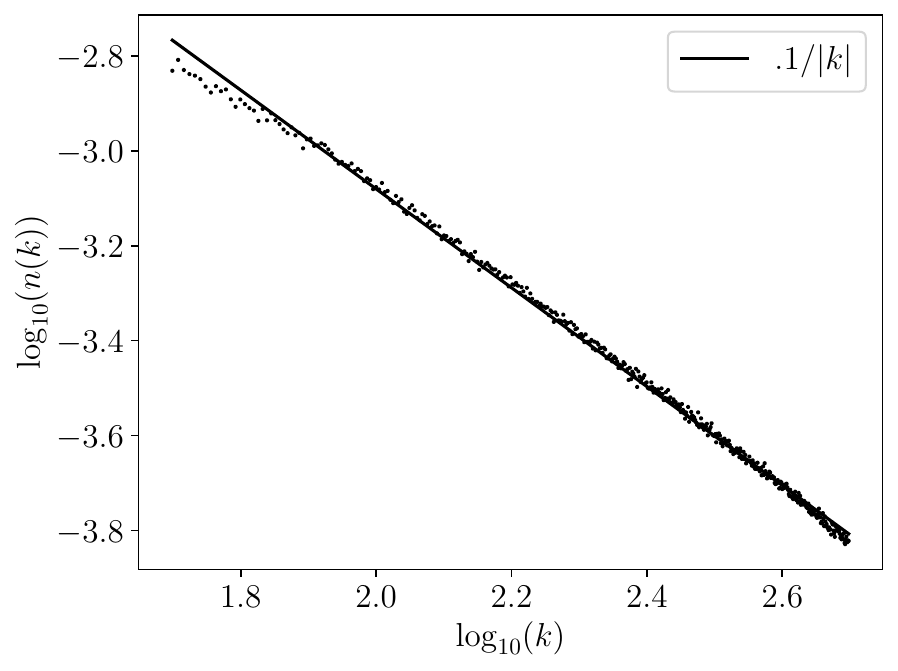}
\caption{Plot of $n_{eq}(k)$ for $k^{i}_{-} < k < k^{i}_{+}$ that we see is  approximated by a fit of $n_{eq}(k)\approx .1/|k|$.}
\label{fig:eqdist}
\end{figure}

Throughout the remainder of this work, we always choose the initial condition 
\[
\hat{\psi}(k,0) = \frac{\epsilon}{|k|}\hat{z}_{k}, ~ \mbox{Re}\left(\hat{z}_{k}\right), \mbox{Im}\left(\hat{z}_{k}\right) \sim \mathcal{N}(0,1)
\]
and parameters 
\[
k^{f}_{-}=6, ~ k^{f}_{+}=9, ~ d_{-}=d_{+}=8,
\]
\[
 k^{d}_{-}=5, ~ k^{d}_{+}=1000, ~ \epsilon=.25.
\]
We take the inertial range to be $k^{i}_{-}<k<k_{+}^{i}$ with $k^{i}_{-}=50$ and $k^{i}_{+}=500$.  We fix the space domain to be $[0, 2\pi]$.  Following the analysis in \cite{majda1}, per our choice of $\epsilon$, the nonlinearity acts over time scales on the order of $1/\epsilon^{2} = 4$ non-dimensional units of time, and we should expect weakly turbulent effects to manifest on time scales on the order of $k^{i}_{+}/\epsilon^{2}$.  Using a pseudo-spectral in space and 4th order Runge--Kutta in time discretization scheme, we use data starting from $t=k^{i}_{+}/\epsilon^{2}$ up to $t_{f}= 101k^{i}_{+}/\epsilon^{2}$, thereby allowing for nonlinearity to induce several turnovers of energy within the inertial range; see Figure \ref{fig:mmtdynamics} for a plot of $|\psi(x,t)|$ for $2k^{i}_{+}/\epsilon^{2} < t < 2k^{i}_{+}/\epsilon^{2} + 160$.  We sample the data at a rate of $\delta_{s} = 40 = 10/\epsilon^{2}$ units of non-dimensional time, corresponding to the timescale of wavenumbers just at the end of the forcing regime.    
Averaging over the timescale $101k_{+}/\epsilon^{2}$, we generate the following approximation of $n_{eq}(k)$ seen in Figure \ref{fig:eqdist}
Thus we see that we are generating dynamics consistent with the time and length scale requirements in WWT theory.  

To characterize the transfer of information across scales, we separate the inertial range into four disjoint intervals, say $\Delta_{n}$, with 
\[
\Delta_{n}=\left[k^{i}_{-}+n\frac{(k^{i}_{+}-k^{i}_{-})}{4}, k^{i}_{-} + (n+1)\frac{(k^{i}_{+}-k^{i}_{-})}{4}\right],
\]
for $n=0,1,2,3$.  Since we are tracking how the mean of $\left|\hat{\psi}(k,t)\right|^{2}$ approaches the WWT distribution, define the fluctuations $F_{n}(t)$ as
\[
F_{n}(t) = \frac{\tilde{F}_{n}(t)}{\sqrt{\left<\tilde{F}^{2}_{n}\right>}},
\]
where
\[
\tilde{F}_{n}(t) = \sum_{k\in \Delta_{n} }\tilde{n}_{W}(k,t),
\]
with
\[
\tilde{n}_{W}(k,t) = \left[\left|\hat{\psi}_{k}\right|^{2}\right]_{W} - \tilde{n}_{eq}(k),
\]
and for generic function $f(k,t)$
\[
\left[f(k,\cdot)\right]_{W}(t) = \frac{1}{W}\int_{t}^{t+W}f(k,\tau)d\tau.
\]
Note, $\left<\tilde{F}_{n}\right>=0$, so that we do not need to include it when scaling by the variance.  Using Equation \eqref{eq:interaction} we find that  
\[
\partial_{t}\tilde{n}_{W} = -2 \sum_{k_{1},k_{2},k_{3}}\text{Im}\left(\left[\hat{\psi}_{1}\hat{\psi}_{2}\hat{\psi}_{3}^{\ast}\hat{\psi}_{k}^{\ast}\right]_{W}\right)\delta({\bf k}_{12,3k}),
\]
showing that $\tilde{n}_{W}(k,t)$ is driven by essentially the same wave-mixing process which leads to the formation of the statistically stationary state we are trying to study.  

\begin{figure}
\includegraphics[width=.5\textwidth]{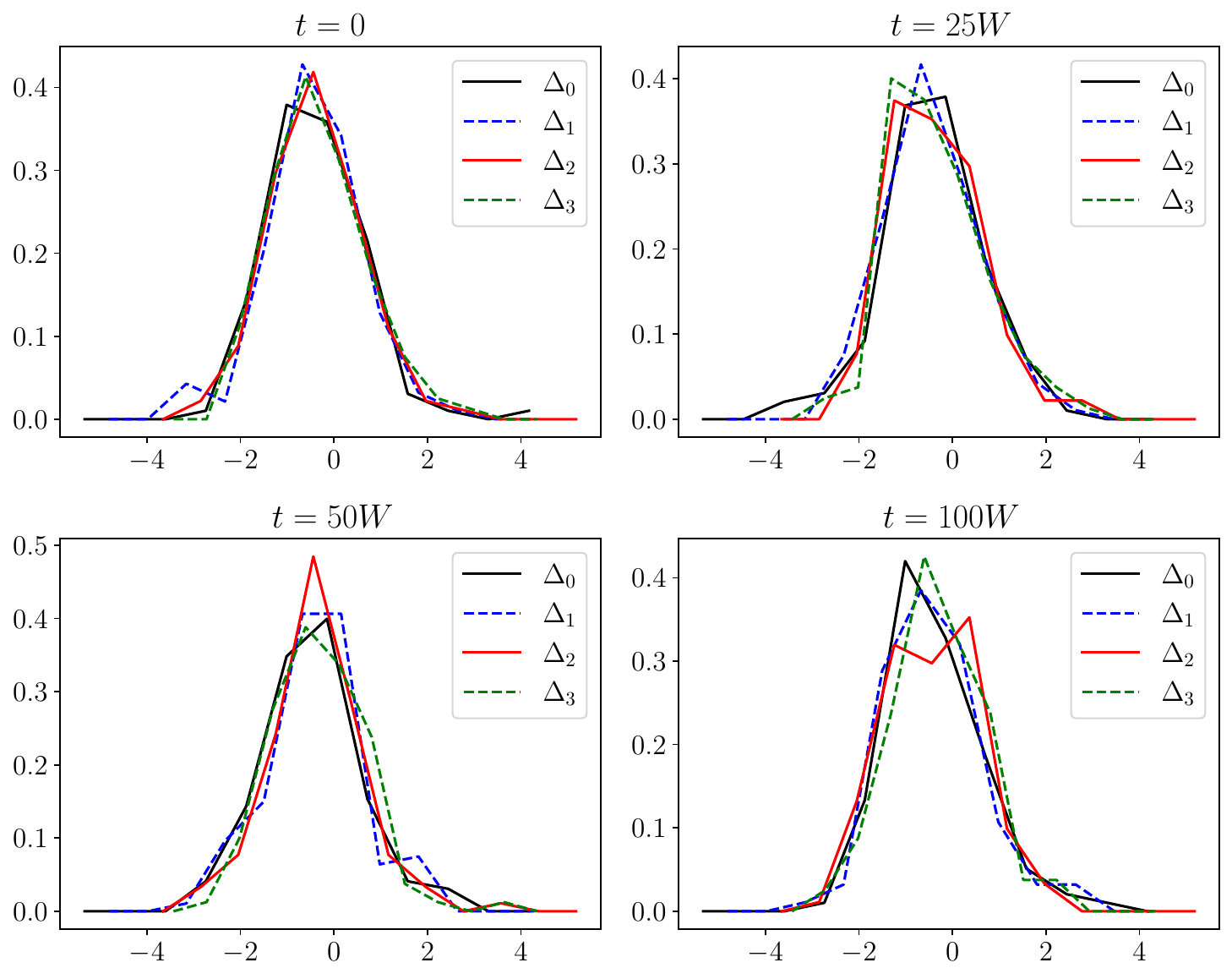}\\
\caption{For window $W=k^{i}_{+}/\epsilon^{2}$, the histograms generated by $\tilde{n}_{W}(k,t)$ over each wavenumber band $\Delta_{n}$.}
\label{fig:time_hist}
\end{figure}

\begin{figure}
\includegraphics[width=.5\textwidth]{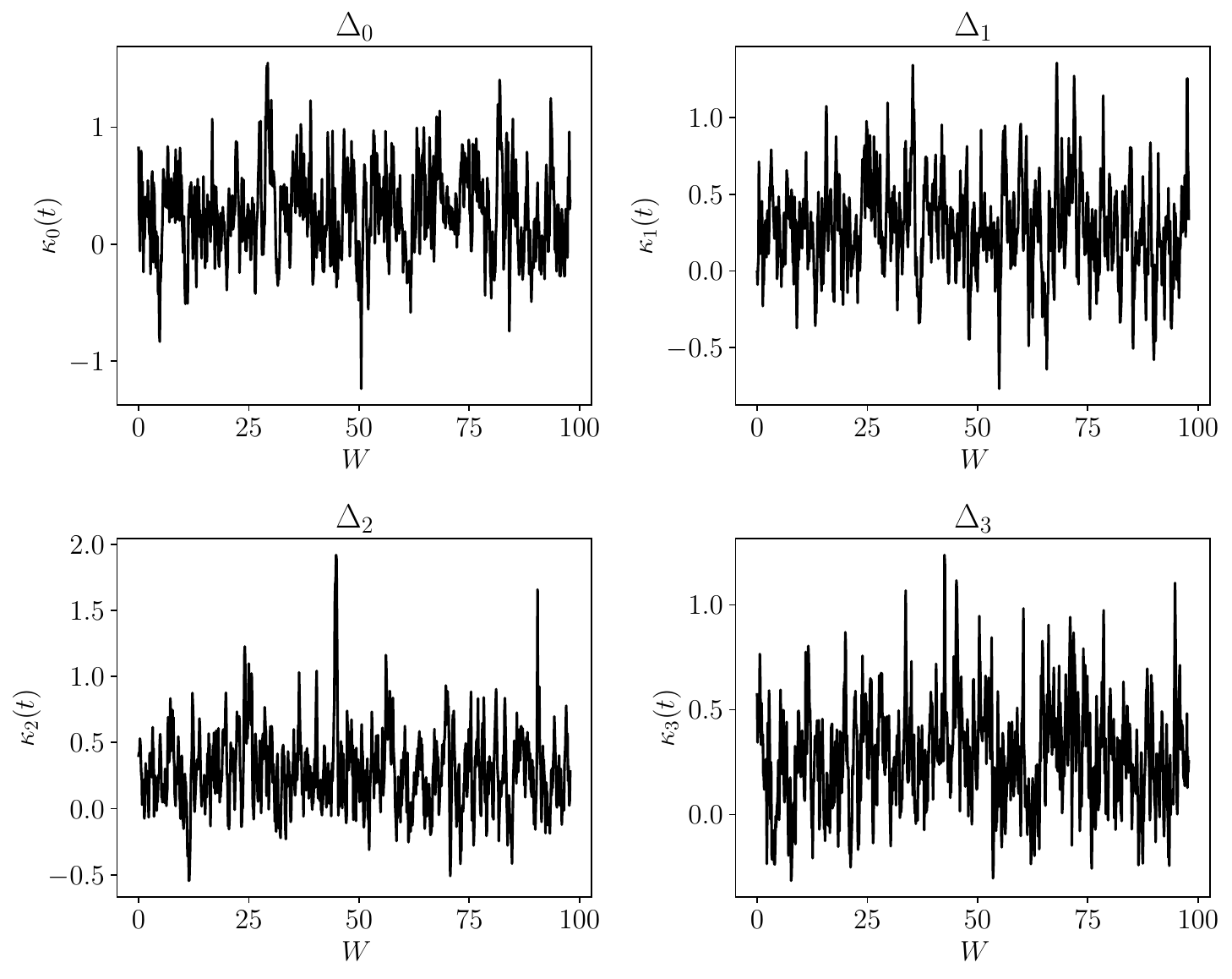}\\
\caption{For window $W=k^{i}_{+}/\epsilon^{2}$, the skewness coefficient $\kappa_{n}(t)$ of the distributions across wavenumber band $\Delta_{n}$ plotted up to $t=100W$.  Each figure is plotted with time-step resolution $W/10=k^{i}_{-}/\epsilon^{2}$.}
\label{fig:skew_hist}
\end{figure}

Choosing $W=k^{i}_{+}/\epsilon^{2}$, we can see how the wave-mixing process drives longer term nonlinear dynamics in $\tilde{n}_{W}(k,t)$ by looking at Figure \ref{fig:time_hist} where we plot the histograms generated by $\tilde{n}_{W}(k,t)$ with $k$ restricted to each wavenumber band $\Delta_{n}$.  Further, in Figure \ref{fig:skew_hist}, we look at the corresponding Fisher--Pearson skewness coefficient $\kappa_{m}(t)$ \cite{NIST_StatHandbook} across each wavenumber band over the same length of time.  Note, in line with our computation of $F_{m}$ we compute each histogram in Figure \ref{fig:time_hist} by normalizing by the mean and standard deviation.  Per the assumptions leading to the derivation of $\tilde{n}_{eq}(k)$ \cite{zakharov}, we see that each distribution is nearly Gaussian, but clearly there are higher moment dynamics representing the active exchange of energy across wavenumbers needed to maintain $\tilde{n}_{eq}(k)$.  This nonlinear exchange is reflected in the oscillations of $\kappa_{n}(t)$ with each distribution over $\Delta_{n}$ exhibiting a positive bias, i.e. $\kappa_{n}(t) > 0$ on average.  This explains the shifting tails and peaks of the distributions seen in Figure \ref{fig:time_hist}.  In turn then, the fluctuations $F_{n}(t)$ track the scaled wavenumber average of the variations seen in said distributions.     

\begin{figure}
\centering
\begin{tabular}{cc}
\includegraphics[width=.23\textwidth]{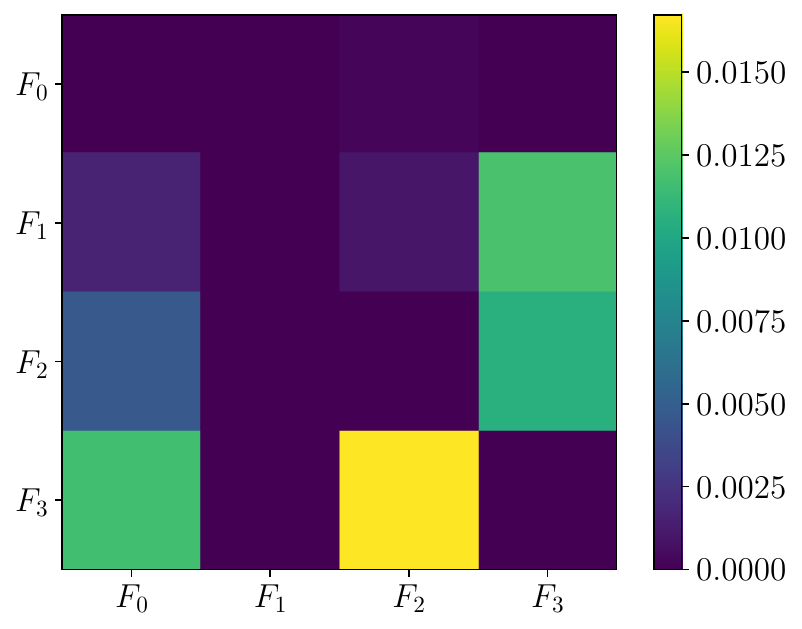} & \includegraphics[width=.23\textwidth]{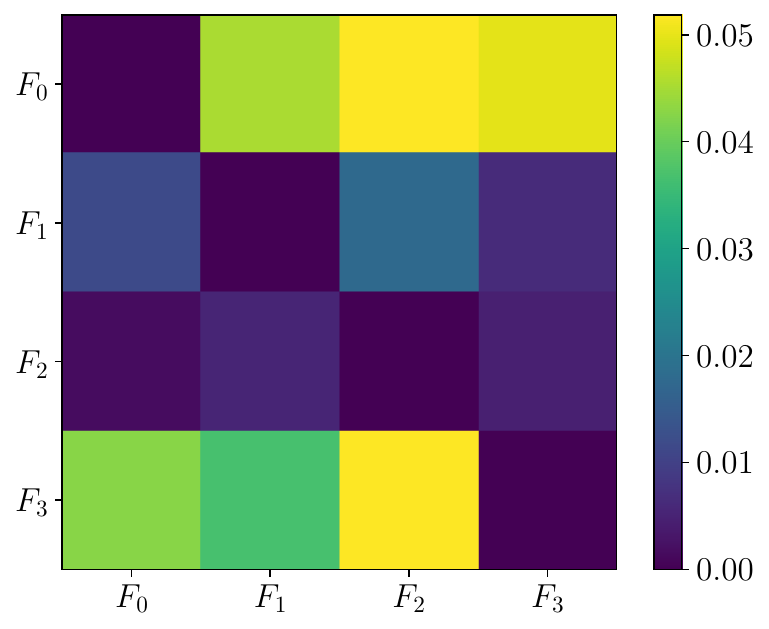}\\
(a) $I^{t,i}$ & (b) $I^{t,m}$\\ 
\includegraphics[width=.23\textwidth]{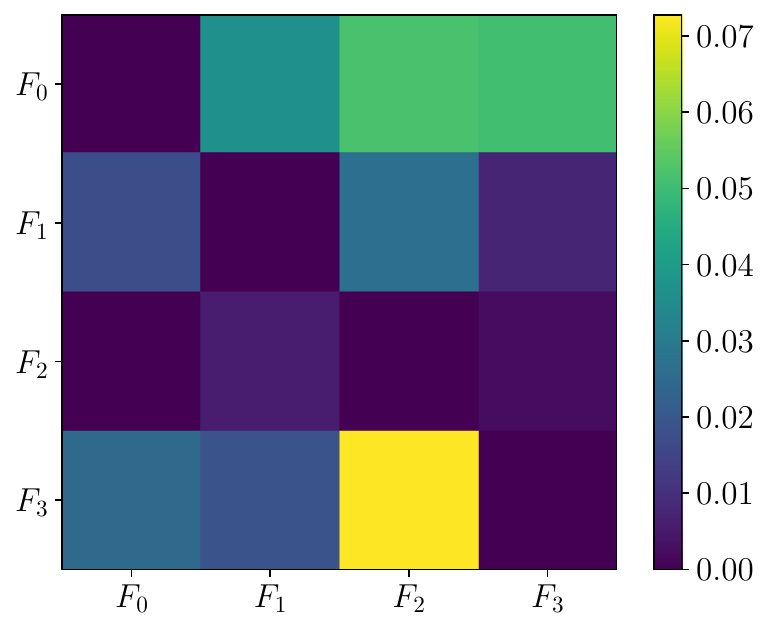} & \includegraphics[width=.23\textwidth]{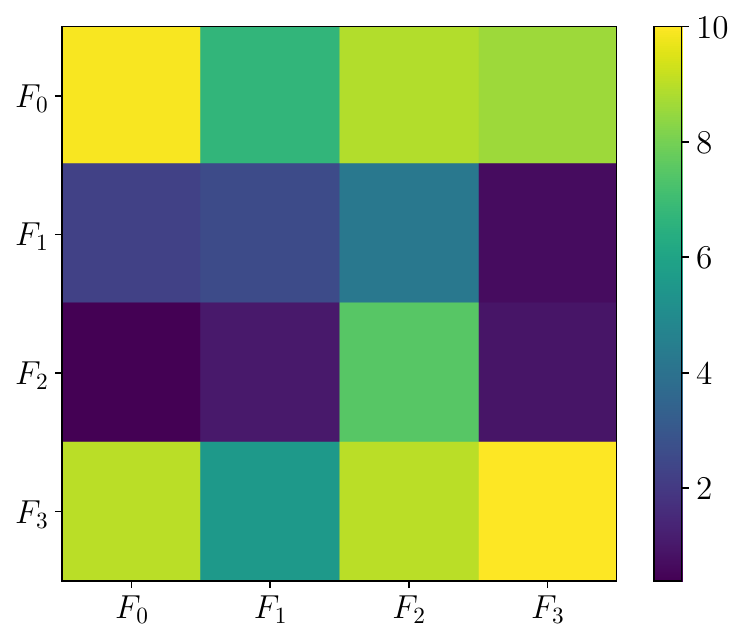}\\
(c) $I^{t,f}$ & (d) Max Lags\\
\includegraphics[width=.23\textwidth]{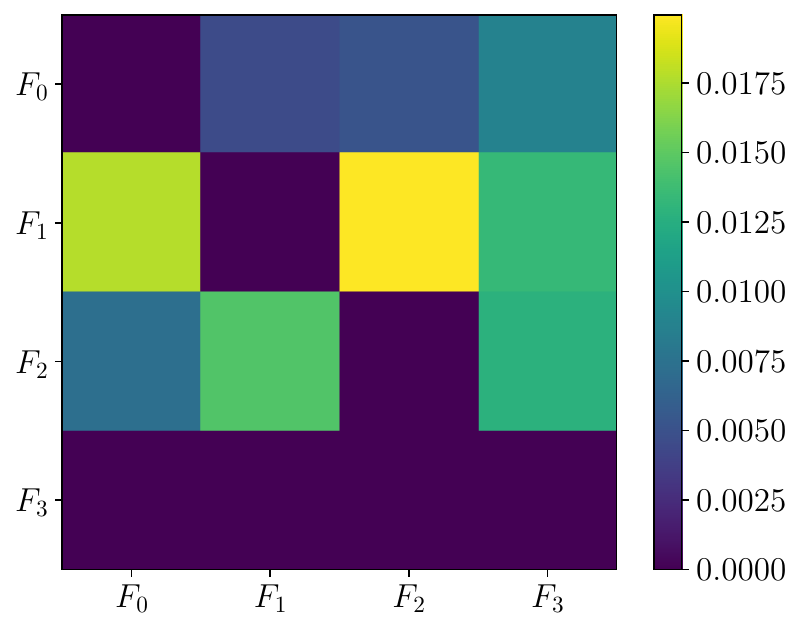} & \includegraphics[width=.23\textwidth]{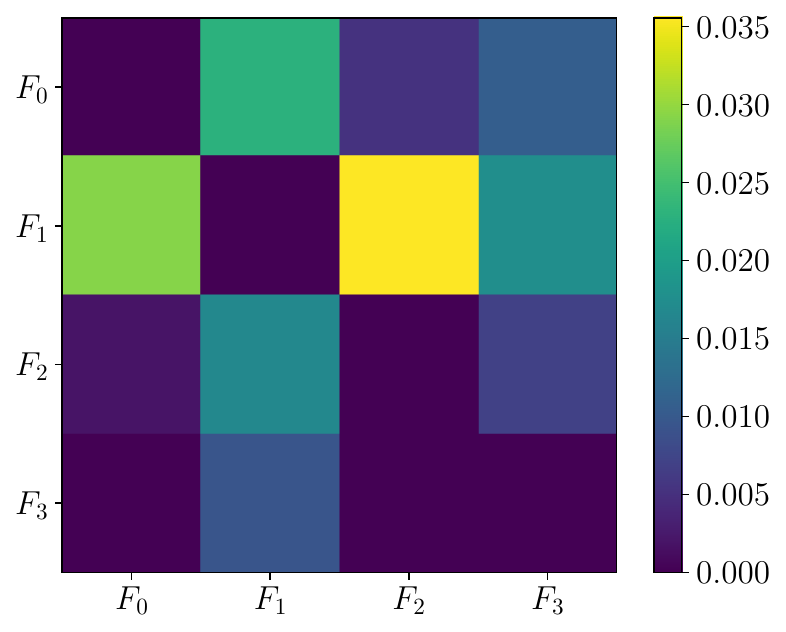}\\
(e) $\sigma(I^{t,m})$ & (f) $\sigma(I^{t,f})$
\end{tabular}
\caption{ The average values over $N_{tr}=100$ trials of (a) $I^{t,i}$, (b) $I^{t,m}$,  (c) $I^{t,f}$, and (d) the maximum lags for the MMT system for $k_{+}/\epsilon^{2}\leq t \leq 101k_{+}/\epsilon^{2}$ averaged over the $W=k^{i}_{+}/\epsilon^{2}$ timescale with $L_{M}=10$ corresponding to one turnover time $W$.  The standard deviations around the averages of $I^{t,m}$ and $I^{t,f}$ are seen in (e) and (f).  The hypothesis-testing threshold is $\alpha_{h}=.05$, the maximum allowed lag $L_{M}=10$ corresponds to a full turnover time $W=k^{i}_{+}/\epsilon^{2}$, and the number of nearest neighbors is $k_{n}=3$.  Note, coupling from target-to-source is read from left-to-right.}
\label{fig:mmt_d_10}
\end{figure}

To analyze the information transfer across the fluctuations $F_{n}(t)$, we first choose the maximal lag $L_{M}=10$, which relative to our rate of sampling, corresponds to one full turnover time $W$ in our data.  Note, a similar approach to that illustrated by the results in Figure \ref{fig:lr_lag_test} was attempted, but no local minimum was seen in $I(F_{n,j+\ell},F_{n,j})$ for any $\ell$.  Thus we choose $L_{M}$ relative to the physics of the MMT equation.    
\begin{figure}
\includegraphics[width=.35\textwidth]{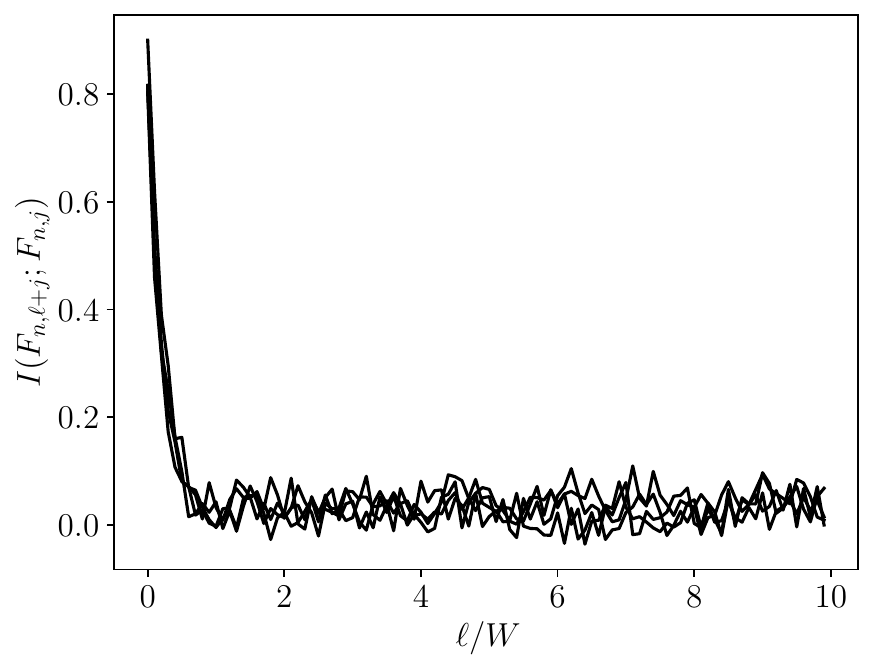}
\caption{$I(F_{n,j+\ell},F_{n,j})$ for $1\leq \ell \leq 10W$.}
\label{fig:scale_info}
\end{figure}
We then run sIDTxl across $N_{tr}=100$ trials and compute the average and variance over $N_{tr}$ of each relevant quantity.  This produces the results of Figure \ref{fig:mmt_d_10}.  Relative to the nearly flat and ambiguous, as regards cascades, decision landscape seen in Figure \ref{fig:mmt_d_10} (a) ($0 \leq I^{t,i}_{nl} \leq .016$), the marginal and final information matrices do not ignore most couplings, but we see a wider range of values in  $I^{t,m}$ and $I^{t,f}$ ($0. \leq I^{t,m}_{nl},I^{t,f}_{nl}  \leq .07$) so that the most prominent couplings are seven times larger than the weakest.  The largest marginal increases in $I^{t,m}$ occur in the forward direction from $F_{2}$ to $F_{3}$ and in the inverse direction from $F_{2}$ to $F_{0}$.  Moreover, the strongest marginal increases are localized in the extremes of the inertial range.  In the forward direction, $F_{0}$, $F_{1}$, and $F_{2}$ as a strong marginal targets relative to the source $F_{3}$.  In the reverse, $F_{0}$ sees the other frequency bands as providing strong marginal target-to-source contributions.  These marginal disparities are somewhat reflected in $I^{t,f}$, but the forward coupling $F_{2}$ to $F_{3}$ is clearly the strongest.  Therefore, we find from these experiments evidence for a preference of the forward over the inverse cascade, but we also see the two processes play essential roles relative to the time scales used in our computations. 

Otherwise, the four-wave resonances driving WWT allow for information transfer in almost all directions, but at markedly lower values of relevance as seen in $I^{t,f}$.  With regards to the lag time scales seen in Figure \ref{fig:mmt_d_10} (d), the largest values seen in $I^{t,f}$ correspond to longer lag choices, reflecting the need for information to propagate across scales over full nonlinear turnover times.  The variances around the means plotted in Figures \ref{fig:mmt_d_10} (e) and (f) show that the larger values in Figures \ref{fig:mmt_d_10} (b) and (c) have the smallest variances and vice versa.  Thus while we can get some degree of variation in the outcomes of the sIDTxl algorithm, it does also tend towards generating clearly distinguishable results of high and low confidence.  

\begin{figure}
\centering
\begin{tabular}{cc}
\includegraphics[width=.23\textwidth]{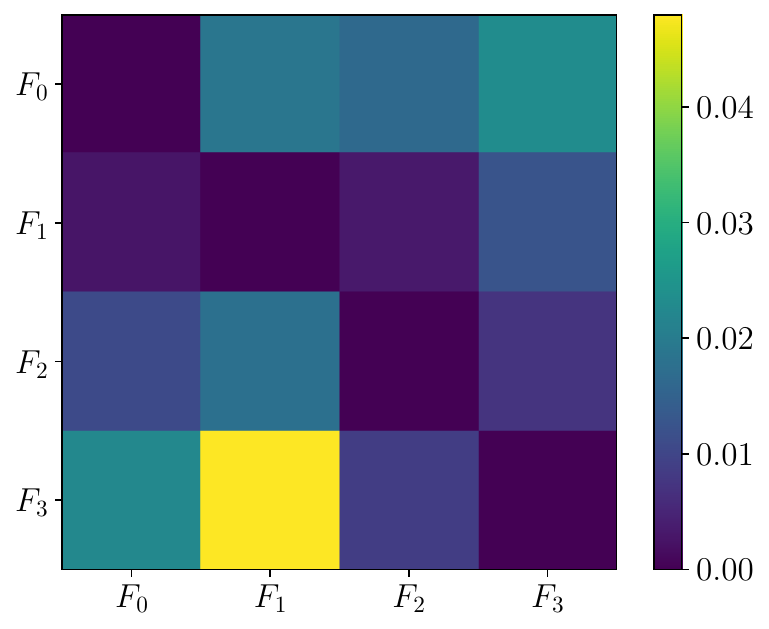} & \includegraphics[width=.23\textwidth]{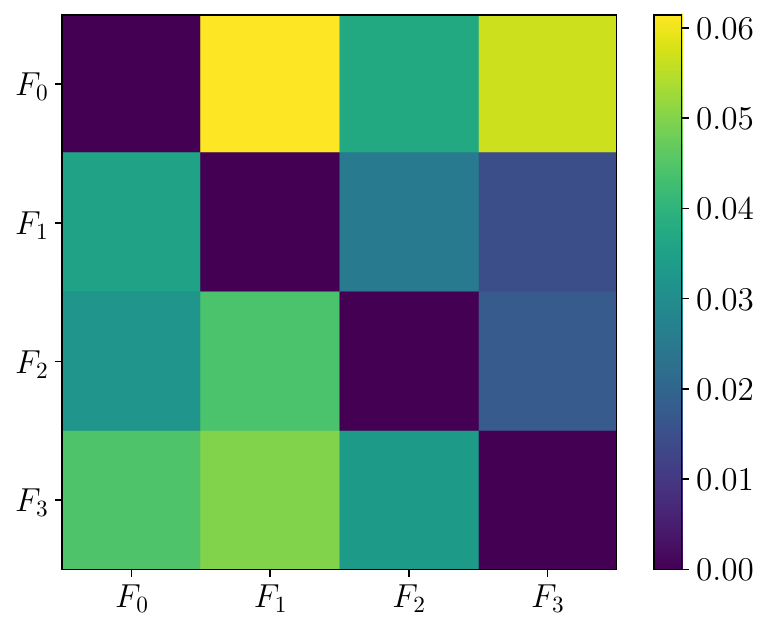}\\
(a) $I^{t,i}$ & (b) $I^{t,m}$\\ 
\includegraphics[width=.23\textwidth]{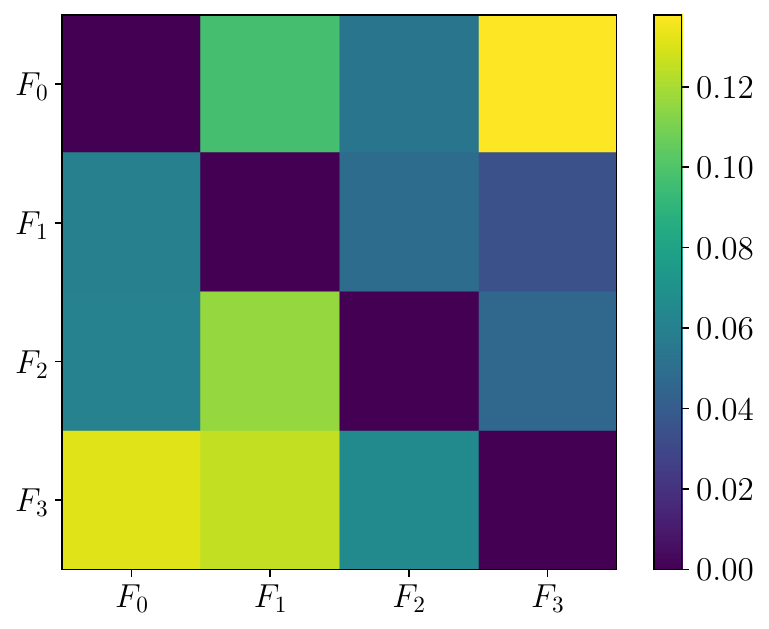} & \includegraphics[width=.23\textwidth]{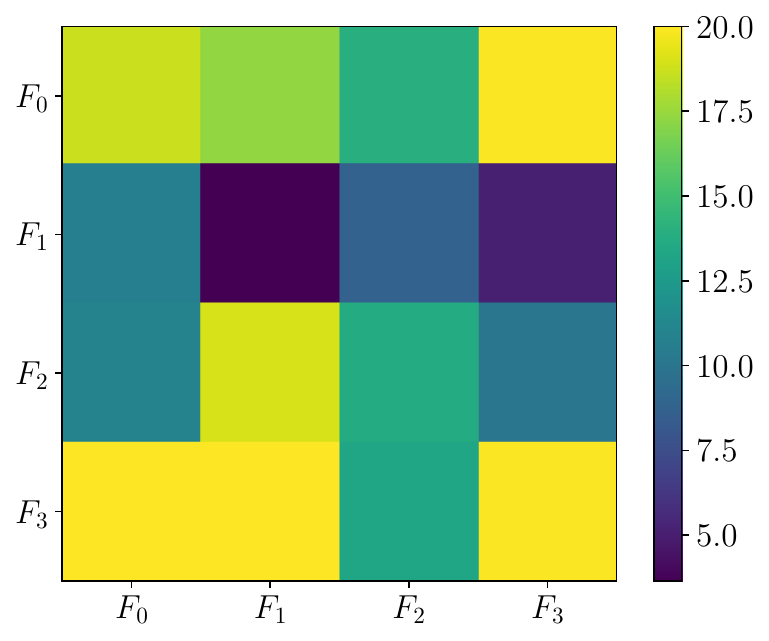}\\
(c) $I^{t,f}$ & (d) Max Lags\\
\includegraphics[width=.23\textwidth]{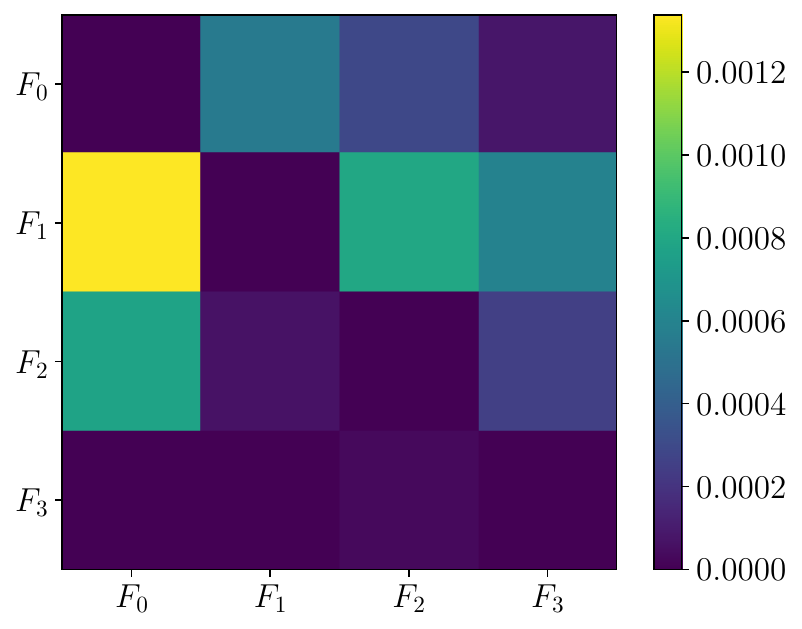} & \includegraphics[width=.23\textwidth]{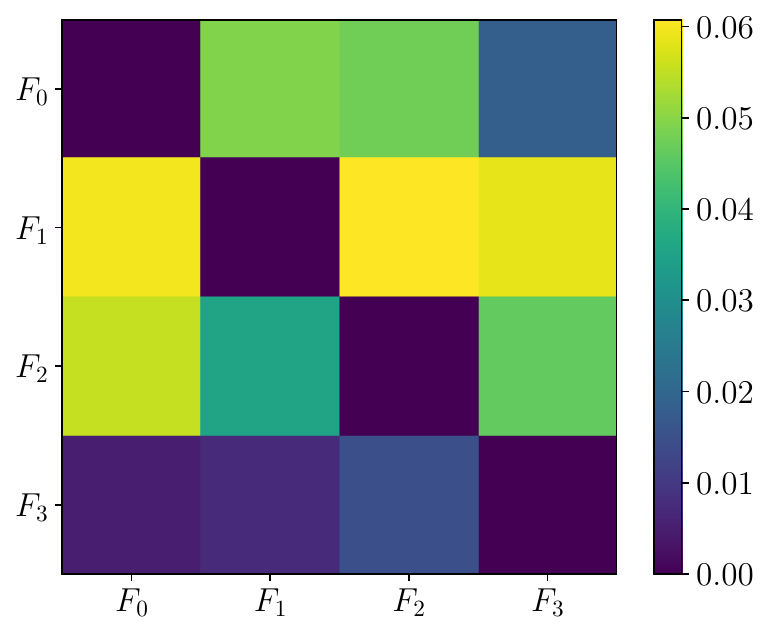}\\
(e) $\sigma(I^{t,m})$ & (f) $\sigma(I^{t,f})$
\end{tabular}
\caption{ Average values over $N_{tr}=100$ trials of (a) $I^{t,i}$, (b) $I^{t,m}$,  (c) $I^{t,f}$, and (d) the maximum lags for the MMT system for $k_{+}/\epsilon^{2}\leq t \leq 101k_{+}/\epsilon^{2}$ averaged over the $W=k^{i}_{+}/\epsilon^{2}$ timescale.  The standard deviations around the averages of $I^{t,m}$ and $I^{t,f}$ are seen in (e) and (f).  The hypothesis-testing threshold is $\alpha_{h}=.05$, the maximum allowed lag $L_{M}=20$ corresponds to two full turnover times $W=k^{i}_{+}/\epsilon^{2}$, and the number of nearest neighbors is $k_{n}=3$.  Note, coupling from target-to-source is read from left-to-right.}
\label{fig:mmt_d_20}
\end{figure}

We now double the maximum lag time so that  $L_{M}=20=2W$.  Looking at Figures \ref{fig:mmt_d_20} (b) and (c), a much wider range of couplings are detected and chosen.  Importantly, the largest values in Figure \ref{fig:mmt_d_20} are almost double those in Figure \ref{fig:mmt_d_10}, showing that fluctuations around the mean in WWT couple and mix across scales over longer time scales than straightforward asymptotic arguments would indicate.  However, comparing Figures \ref{fig:mmt_d_20} (b) and (e) show that some of the mid-value range selections made by sIDTxl also correspond to some of the highest variances, and as before then, they should be considered as less meaningful than low-variance selections.  This also holds true when comparing final target-to-source values as seen in contrasting Figures \ref{fig:mmt_d_20} (c) and (f).  Keeping this in mind then, we again find more information flowing in the direct cascade direction, though again, by allowing longer maximum lag times, we get a richer picture of multiscale interactions.  Likewise, comparing to the $L_{M}=10$ results, we see a degree of consistency in which targets couple to which sources most strongly, so we get some notion of convergence by extending the maximum lag.  However, referring to Figures \ref{fig:mmt_d_20} (b) and (d), most of the selections made in $I^{t,m}$ use longer lags compared both within this case but also to the $L_{M}=10$ case.       

\section{Discussion and Future Work}
In all then, we see that our approach provides a sophisticated lens through which to view the dynamic equilibrium sustaining $\tilde{n}_{eq}(k)$ while it is actively forced and damped.  In all cases, while a bias for the forward cascade of energy being more dominant is present, we also see that the four-wave mixing process driving dynamics couples scales in inverse directions as well.  Thus our method provides a complex and dynamic landscape characterizing weak turbulence.  As we know from \cite{lvov} though, the wave mixing driving energy transfer prevents the formation of as straightforward cascades of information as we would expect in the case of viscously damped three dimensional turbulence \cite{lozano}. 
Thus our approach provides a relatively condensed but statistically meaningful way to characterize this intricate balance.  

Given the relative success of our method and its ability to describe how nonlinear waves interact across disparate time scales and wavenumbers, we have developed a useful tool for data-driven analysis. We anticipate that our technique will enable us to uncover relationships between variables at different time scales, helping us analyze how systems evolve as parameters change. In particular, we know that certain systems, including the MMT equation, undergo critical events—such as the onset of extreme behavior—that exhibit observable precursors. These precursors may manifest as shifts in interaction time scales, either becoming critical or obstructed. We hope that our approach will provide a bifurcation analysis of criticality across time scales, offering insights that can be connected to more traditional methods.

\section{Acknowledgments}
We are grateful to the support from Office of Naval Research grant N00014-23-1-2106.  Also, E.M.B. is supported by  the NIH-CRCNS, DARPA RSDN, the ARO, the AFSOR and the ONR.  We are likewise grateful to Daniel Jay Alford-Lago and Stefan Cline for many fruitful discussions and suggestions that have helped improve this work.  

\bibliographystyle{unsrt}
\bibliography{ionosphere_bib}
\end{document}